\documentclass[11pt]{article}
\pdfoutput=1

\usepackage{jheppub}
\usepackage{rotating}
\usepackage{enumitem}
\usepackage{wrapfig}
\usepackage{comment}

\usepackage{amsfonts}
\usepackage{amsmath}
\usepackage{slashed}
\usepackage{amssymb}
\usepackage{latexsym}
\usepackage{multirow}
\usepackage{hyperref}
\usepackage{enumitem}
\usepackage{mathtools}
\hypersetup{colorlinks=true,citecolor=red,linkcolor=magenta,urlcolor=blue}
\usepackage{relsize}
\usepackage{booktabs}
\usepackage{subfigure}
\usepackage{cleveref}
\usepackage{fancyhdr}
\usepackage{float}
\usepackage{graphicx}
\usepackage{microtype}
\usepackage{tabularx}
\usepackage{xcolor}
\usepackage{orcidlink}
\usepackage{siunitx,adjustbox,bm}
\usepackage[bottom]{footmisc}
\usepackage{subfigure}

\usepackage{tikz}
\usetikzlibrary{positioning, arrows.meta, calc, fit, backgrounds}
\usepackage{longtable}

\newcommand{\mttbar}{m(t\bar{t})}
\renewcommand{\mid}{|}

\title{
\parbox{0.6\textwidth}{
\centering
Big Dipper, Help Me Find A Way\\
{\large --- Dip-hunting at hadron colliders ---}}
\hfill
\parbox{0.3\textwidth}{\vspace{-3.5cm}\includegraphics[width=0.3\textwidth]{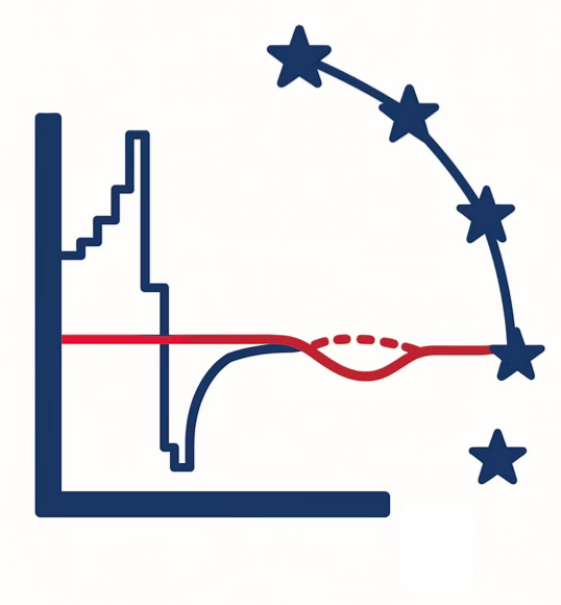}}
}
\author[a]{Diego A. Baron Moreno\orcidlink{0000-0001-9864-7985},}
\author[a]{Christoph Englert\orcidlink{0000-0003-2201-0667},} 
\author[a]{Yvonne Peters\orcidlink{0000-0003-1702-7544}}
\affiliation[a]{Department of Physics \& Astronomy, University of Manchester,\\Oxford Road, Manchester M13 9PL, United Kingdom}
\emailAdd{diego.baronmoreno@manchester.ac.uk}
\emailAdd{christoph.englert@manchester.ac.uk}
\emailAdd{yvonne.peters@manchester.ac.uk}
\abstract{Destructive interference between signal and background processes poses a fundamental challenge in searches for top-philic scalar resonances, significantly reducing experimental sensitivity to well-motivated extensions of the Higgs sector. Traditional bump-hunting strategies fail in this instance because interference effects invalidate the narrow-width approximation across large regions of the BSM parameter space. As a result, experimental analyses typically rely on detailed simulations to accurately model these effects throughout the full analysis chain. In this work, we consider the inverse problem in a proof-of-principle study: given an observed pattern in a discriminating distribution, what is the likelihood that it originates from a BSM scalar? To address this, we employ parametric neural networks to learn the likelihood ratio as a function of both background and key BSM parameters, based on a ratio-of-signed-mixtures framework. We perform inference by testing the compatibility of observed data with a scan over the parameter space of a minimal scalar extension of the Standard Model. While BSM parameter extraction remains inherently model-dependent, our approach provides a robust diagnostic in perturbative regimes and motivates a complementary strategy of `dip-hunting'. This strategy extends traditional bump-hunts and could point the way as we navigate towards future discoveries.} 
\begin{document}
\maketitle
\allowdisplaybreaks
\flushbottom
%%%%%%%%%%%%%%%%%%%%%%%%%%%%%%%%%%%%%%%
\section{Introduction}
\label{sec:intro} 
%%%%%%%%%%%%%%%%%%%%%%%%%%%%%%%%%%%%%%%
Searches for resonantly produced new physics beyond the Standard Model have, so far, not revealed convincing evidence for the presence of new particles. This outcome is puzzling, as the electroweak scale is a well-motivated interface bridging the Standard Model (SM) to a more complete theory of particle interactions, addressing the SM's shortcomings. Well-known ways to `hide' the presence of scalar exotics include accidental signal-background interference from new physics that is top-philic~\cite{Gaemers:1984sj}. This includes a wide range of SM Higgs-sector extensions~\cite{Dicus:1994bm,Basler:2019nas}, which are under active investigation by the experiments at the Large Hadron Collider (LHC). Any such signal-background interference, if experimentally verified, would be a strong probe of the underlying model dynamics~\cite{Jung:2015gta,Frederix:2007gi,Djouadi:2019cbm,Carena:2016npr}. Furthermore, recent analyses of concrete new physics scenarios that enable a cross-correlation of Higgs signal strength measurements, flavour constraints, and direct search limits point towards an overall destructive interference pattern when the mass resolution of the $t\bar t$ system is realistic. Sensitivity to the emerging dip-structures is limited by systematic uncertainties, see the recent Refs.~\cite{ATLAS:2024vxm,CMS:2025dzq}. Framing of a local excess in an invariant mass distribution is quantitatively possible in the narrow-width approximation when interference is suppressed. Conversely, when interference is significant, the narrow-width approximation breaks down irretrievably. 

A practical question, therefore, arises: Whilst a local excess can be translated into the presence of a new particle, a local deficiency in data does not directly admit such an interpretation. In this work we address the question of how an observed destructive interference pattern is consistent with a particle interpretation given an expected background hypothesis. This question is not too dissimilar from the traditional approach of correlating a model with a specific final-state analysis, while tracking potential interference effects along the way. `Hunting' such a dip in data is the focus of this work. In parallel, our approach has the potential of opening up a more economical avenue for experimental interpretations.

As background distributions are typically difficult to model from first principles, in bump-hunting, they are constrained using a range of data techniques~\cite{CMS:2020zti,CMS:2019gwf}. Extending this to negative weights that interference-corrected signals are characterised by, we motivate `dip-hunting' to help us find the way to BSM discovery. This paper is organised as follows: In Sec.~\ref{sec:intf}, we review the interference effects that are sensitivity-limiting in $t\bar t $ resonance searches. We also detail our event simulation setup and cross-checks. Section~\ref{sec:methodology} introduces an approach for dip-hunting using a Ratio of Signed Mixtures model~\cite{Drnevich:2024vfj}; we also detail the network architecture that is used to extract the $t\bar t$ dip structure. Section~\ref{sec:res} is devoted to results, where we show in a proof-of-principle investigation that dip-hunting becomes possible with the procedure laid out in Sec.~\ref{sec:methodology}. We also comment on model dependencies and the architecture's robustness, providing pointers for future studies. We conclude in Sec.~\ref{sec:conc}.

%%%%%%%%%%%%%%%%%%%%%%%%%%%%%%%%%%%%%%%%%%%%%%
\section{Interference in resonant top pair production}
\label{sec:intf}
%%%%%%%%%%%%%%%%%%%%%%%%%%%%%%%%%%%%%%%%%%%%%%
Interference in $gg \to t\bar t$ arises when a significant coupling for an exotic scalar $S$ to top quarks is present. This enables the production via gluon fusion, $gg\to S$, and the subsequent decay, $S\to t\bar t$. Such signatures directly arise in non-singlet Higgs sector extensions of the SM: The combination of Higgs fields orthogonal to the iso-triplet giving rise to the $W, Z$ gauge boson masses is naturally gauge-phobic. Once the exotic scalar is sufficiently heavy, $t\bar t$ final states typically dominate the electrically neutral exotics' decay phenomenology. Concrete extensions, such as the Two-Higgs-Doublet Model, are therefore particularly susceptible to interference effects, especially after the first Higgs investigations force the 125 GeV Higgs boson towards the alignment limit~\cite{Basler:2019nas}. When these effects are included in the existing experimental searches, they give rise to significant systematic uncertainties~\cite{ATLAS:2025kmo}. From a theoretical point of view, indirect EFT-based analyses of di-top final states are unlikely to obtain competitive constraints to such scenarios given current QCD uncertainty expectations~\cite{BessidskaiaBylund:2016jvp,Englert:2019rga}.

In the following, we will parametrise the exotic scalar through a simplified model
\begin{equation}
\label{eq:model}
{\cal{L}} = {\cal{L}}_{\text{SM}} - C_{e} \, \frac{y_t}{\sqrt{2}} \, S \, \bar t t   + {1\over 2}(\partial_\mu S)^2-{m_S^2\over 2} S^2\ ,
\end{equation}
normalising to the SM's top quark expectation (via the Yukawa $y_t$) as typically performed in related experimental analyses. The decay width $S\to t\bar t$ (denoted in the following by $\Gamma_S$) and one-loop $g g\to S$ amplitude are widely documented in the literature, e.g.~\cite{Georgi:1977gs,Djouadi:2005gi,Plehn:2009nd}. The decay width is dominantly described by
\begin{equation}
\label{eq:4.1}
    \Gamma_S = {3C_{e}^2 m_t^2\over 8\pi v^2} {m_S} \left({{1-{4m_{t}^2 \over m_S^2}}}\right)^{3/2} \ , 
\end{equation}
with $v\simeq 246~\text{GeV}$ and $m_t\simeq 172~\text{GeV}$.\footnote{Ultraviolet finite loop-induced decays $S\to gg,\gamma\gamma,\gamma Z$ increase this decay width at the 1\% level. As this does not impact our results, we ignore this small numerical effect in the following.} We will limit ourselves to CP-even couplings, but we will 
%%%%%%%%%%%%%%%%%%%%%%%
\begin{wrapfigure}[11]{l}{0.5\textwidth}
\centering
\parbox{0.5\textwidth}{
\includegraphics[width=0.48\textwidth]{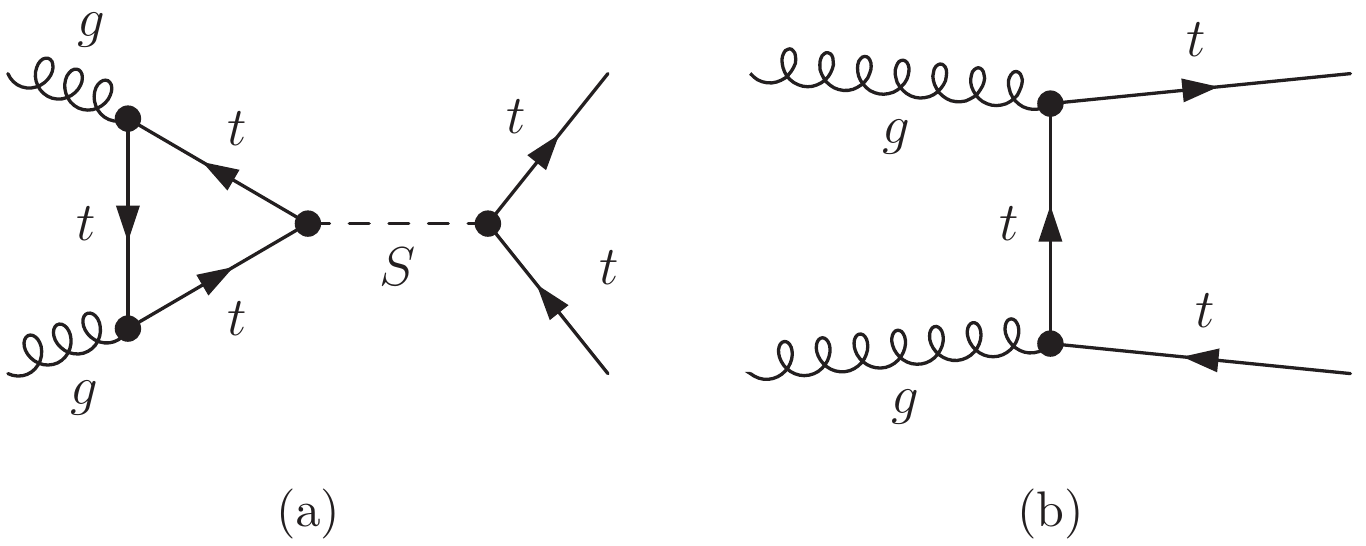}}
\caption{\label{fig:feyn} Representative Feynman diagrams contributing to signal ($gg\to S\to t\bar t$)-background ($gg\to t\bar t$) interference at leading order.}
\end{wrapfigure}
%%%%%%%%%%%%%%%%%%%%%%%
come back to the broader question of model-dependence in Sec.~\ref{sec:modeldependence}. For perturbative choices of the couplings $C_e\sim 1$, we typically are in a parameter region $\Gamma_S/m_S\lesssim 10\%$ where the phenomenology is shaped by a characteristic dip structure (see, e.g.,~\cite{ATLAS:2025kmo}), which is our main motivation to focus on the simple extension as a test case.

The signal $g g\to S \to t\bar t$ amplitude ${\cal{M}}_{S}$ then interferes with the continuum QCD background $g g\to t \bar t$ via the latter's $t$-channel virtual top exchange diagram, which contains a colour-singlet component (we denote this by ${\cal{M}}_{\text{con.},t}$). Focusing on this trivial colour structure alone, the interference arises as (see Fig.~\ref{fig:feyn})
\begin{equation}
\label{eq:interf}
|{\cal{M}}|^2 = |{\cal{M}}_{S}|^2 + 2\, \text{Re} \left( {\cal{M}}_{S}^\ast\, {\cal{M}}_{\text{con.},t} \right)+ |{\cal{M}}_{\text{con.},t}|^2 + \dots \ , 
\end{equation}
where the ellipses contain other background-relevant amplitude pieces, i.e. the colour non-singlet $t-$ and $s-$channel QCD continuum contributions. Modelling the scalar's width as a Breit-Wigner distribution is phenomenologically admissible in the perturbative regime $\Gamma_S \ll m_S$~\cite{Goria:2011wa, Seymour:1995np, Papavassiliou:1996zn, Papavassiliou:1997fn, Englert:2015zra}. As the `background' is predominantly a smooth function in this kinematic region, this leads to large distortions of the typically narrow $S$ line shape, not usually resolvable through the detectors' finite resolution in concrete scenarios like the Higgs doublet models~\cite{Basler:2019nas}.  For scenarios that can be described by Eq.~\eqref{eq:interf}, i.e. situations where we can trust the perturbative Monte Carlos workflow that underpins the experimental analysis of the hard scattering region, it is clear that there is analytical, albeit model-dependent, control over the interference region. After all, interference effects are explicitly accounted for in the existing experimental analysis strategies. Any background-only compatible distribution will always be characterised by local statistical variations, and injecting a signal-only hypothesis can be relatively straightforwardly phrased as a local $p$ value to check for the presence of new interactions. A generalisation of this question is then how local fluctuations in data can point towards the interference-obscured presence of new physics once the background hypothesis is assumed to be known (and the signal hypothesis is characterised by relatively few phenomenologically relevant parameters). This is the focus of the presented paper.

To include the interference effects, we map the simplified model of Eq.~\eqref{eq:interf} onto an effective implementation of $gg\to S(q^2)$, keeping the top mass dependence at one loop and full off-shell dependence $q^2\neq 0$. The momentum-dependent form factor can then be included in the dedicated {\texttt{Helas}}~\cite{Murayama:1992gi} routine generated as part of a {\texttt{UFO}}~\cite{Degrande:2011ua, Darme:2023jdn} interface with {\texttt{MadGraph5\_aMC@NLO-v3.5.3}}~\cite{Alwall:2014hca}. It is then straightforward to model interference-corrected cross sections directly with {\texttt{MadGraph5\_aMC@NLO-v3.5.3}}. We have cross-checked our implementation analytically~\cite{Kniehl:1995tn, Djouadi:2005gi, Plehn:2009nd} as well as numerically against Ref.~\cite{ATLAS:2025kmo}. Throughout, we focus on 13 TeV LHC collisions.

%%%%%%%%%%%%%%%%%%%%%%%%%%%%%%%%%%%%%%%%%%%%%%
\section{Dip-Hunting Methodology}
\label{sec:methodology}
%%%%%%%%%%%%%%%%%%%%%%%%%%%%%%%%%%%%%%%%%%%%%%
At the experimental level, we have access to observables that are collected event-by-event into histograms that allow for statistical interpretations. Observable distributions are a reflection of the underlying probability density generated by the parametric physics model $p(\boldsymbol{x}|\boldsymbol{\theta})$, where $\boldsymbol{x}$ represents the event observables, and $\boldsymbol{\theta}$ the parameter set governing the model. In the following, we will use Machine Learning (ML) methods and Monte Carlo (MC) simulations to compute $p(\boldsymbol{x}|\boldsymbol{\theta})$. Our method relies on mapping from the probability density of a reference MC sample, in this case the SM (background) $t\bar{t}$ density ($p(\boldsymbol{x},\theta_{\text{ref}})$), to the general $p(\boldsymbol{x}|\boldsymbol{\theta})$ density. In the inference step, we derive the set of physics parameters ($\hat{\boldsymbol{\theta}}$) that are most compatible with hypothetically observed data. The inference step is based on a scan over the model space $\{\boldsymbol{\theta}\}$ via the $p(\boldsymbol{x}|\boldsymbol{\theta})$ distribution, which is obtained from our reference to target mapping, contrasting the result of the mapping with the hypothetically observed data $\bold{X_{\text{hyp}}}=\{x_{\text{hyp}}\}_{i=1}^{N}$, where $N$ is the number of hypothetical data events.

%%%%%%%%%%%%%%%%%%%%%%%%%%%%%%%%%%%%%%%%%%%%%%
\subsection{The Ratio of Signed Mixtures Model}
\label{sec:rosmm}
%%%%%%%%%%%%%%%%%%%%%%%%%%%%%%%%%%%%%%%%%%%%%%
Due to the interference effects of Eq.~\eqref{eq:interf}, the $p(\boldsymbol{x}\mid\boldsymbol{\theta})$ distribution is generally a quasi-probabilistic density. This means that for some regions in the $\boldsymbol{x},\boldsymbol{\theta}$ space, $p(\boldsymbol{x}|\boldsymbol{\theta})$ will take negative values. A well-established method to estimate the quasi-probabilistic likelihood ratio (qLLHR) between two quasi-probabilistic distributions,
\begin{equation}
    r(\boldsymbol{x}, \vec{c}\,) = \frac{p(\boldsymbol{x})_{\text{target}}}{p(\boldsymbol{x})_{\text{ref}}},
\end{equation} 
is the Ratio of Signed Mixtures Model (RoSMM)~\cite{Drnevich:2024vfj}. The function $r(\boldsymbol{x},\vec{c}\,)$ maps between the reference and the parametrised distributions. The $\vec{c}$ coefficients need to be determined as part of the RoSMM method. Their derivation will be explained in this section. 

In the RoSMM method, multiple classifiers are trained to separate the \textit{reference} class (0-class: the SM events in this case), and the \textit{target} class (1-class: the BSM events generated from the model described in Sec.~\ref{sec:intf}). The MC events get split according to the sign of the MC weight. This generates four sub-samples
\begin{equation}
    \mathcal{D}_{Y_i}(w_{\pm}) \leftarrow 
    \begin{cases*}
    \{(\boldsymbol{x},w) : (\boldsymbol{x},w) \in \mathcal{D}_{Y_i}, w \ge 0\}\\
    \{(\boldsymbol{x},|w|) : (\boldsymbol{x},w) \in \mathcal{D}_{Y_i}, w < 0\}
    \end{cases*},
    \label{eq:3.2}
\end{equation}
where $Y_i=0,1$ indicates the event class. According to the RoSMM method, the qLLHR is decomposed into four LLHR sub-density ratio contributions:
\begin{equation}
    r(\boldsymbol{x}, \vec{c}\,) = \left [ \left(\frac{c_0}{c_1}\right)r_{++}^{-1}(\boldsymbol{x}) + \left(\frac{1 - c_0}{c_1}\right) r_{-+}^{-1}(\boldsymbol{x}) \right ]^{-1} + \left [ \left(\frac{c_0}{1 - c_1}\right)r_{+-}^{-1}(\boldsymbol{x}) + \left(\frac{1 - c_0}{1 - c_1}\right) r_{--}^{-1}(\boldsymbol{x}) \right ]^{-1},
    \label{eq:3.3}
\end{equation}
where 
\begin{equation}
    r_{+-}(\boldsymbol{x}) \equiv \frac{p_{w_{-}}(\boldsymbol{x} | Y = 1)}{p_{w_{+}}(\boldsymbol{x} | Y = 0)},
\end{equation}
and equivalently for $r_{\pm\pm}(\boldsymbol{x})$. The $\vec{c}=\{c_0,c_1\}$ coefficients are estimated via
\begin{equation}
    \hat{c}_{Y} = \frac{\sum_{i,w_i\ge 0} w_i}{\sum_{i}w_i},
\end{equation}
where $w_i$ are the MC events weights that belong to the dataset with label $Y$. One classifier is trained for each sub-density ratio $r_{\pm\pm}(\boldsymbol{x})$ using the appropriate dataset as defined in Eq.~\eqref{eq:3.2}.

In our case, all the SM events have positive MC weights. This implies that $\hat{c}_0 = 1$, and it also simplifies Eq.~\eqref{eq:3.3} to
\begin{equation}
    r(\boldsymbol{x}, \vec{c}\,) = c_1 r_{++}(\boldsymbol{x}) + (1 - c_1)r_{+-}(\boldsymbol{x}).
    \label{eq:3.6}
\end{equation}
The goal is to learn $r(\boldsymbol{x},\vec{c}\,)$ as a function of the BSM parameters $\boldsymbol{\theta}$. To achieve this, we generate several BSM samples, each with the same number of events but with different values of the $\boldsymbol{\theta}$ parameters that govern the new physics. In our case, we vary the resonance mass $m_S$ and the coupling $C_e$. Our class-1 dataset is formed by combining all datasets generated with different parameter choices. To include the $\boldsymbol{\theta}$ dependence in the ML sub-density estimators, we feed the event parameters alongside the event observables as input to the neural network (NN) classifiers. 

%%%%%%%%%%%%%%%%%%%%%%%%%%%%%%%%%%%%%%%%%%%%%%
\subsection{Neural Network Architecture}
\label{sec:arch}
%%%%%%%%%%%%%%%%%%%%%%%%%%%%%%%%%%%%%%%%%%%%%%
The input data consists of simulated events containing parton-level top-quark kinematic variables and global event quantities\footnote{Unfolding to top quark level has been successfully demonstrated throughout the LHC top quark programme, see e.g.~\cite{ATLAS:2016bac,ATLAS:2022mlu}, and has been used in a variety of follow-up studies, e.g.~\cite{Buckley:2015lku,Brivio:2019ius}.}, in addition to one or more theory parameters $\boldsymbol{\theta}$ that characterise the signal hypothesis. Table~\ref{tab:observables} summarises the observables used in the
different analysis configurations: 1D configurations only use either the coupling or mass as the inference target, and 2D configurations have both parameters as the inference objective.\footnote{At parton-level, the kinematics are described by two parameters which can be chosen as the centre-of-mass energy and the scattering angle. The chosen kinematical variables are therefore correlated, but more kinematic information becomes relevant when considering the impact of decay, showering, and hadronisation, which will be studied elsewhere~\cite{forth}.} The 1D and 2D configurations allow us to study the problem with increasing complexity, providing a better understanding of the methodology limitations. 
%
%%%%%%%%%%%%%%%%%%%%%%%%%%%%%%%%%%%%%%%%%%%%%%
\begin{table}[!t]
  \centering
  \begin{tabular}{llll}
    \toprule
    \textbf{Variable} & \textbf{Description} & \textbf{Transformation} & \textbf{Configurations} \\
    \midrule
    $p_{\mathrm{T}}^{t_1}$
      & Leading top-quark transverse momentum
      & $\log(x + 10^{-3})$
      & 1D, 2D \\
    $\eta^{t_1}$ 
      & Leading top-quark pseudorapidity
      & Standard scaling
      & 1D, 2D \\
    $\eta^{t_2}$ 
      & Sub-leading top-quark pseudorapidity
      & Standard scaling
      & 1D, 2D \\
    $m_{t\bar{t}}$ 
      & Invariant mass of the $t\bar{t}$ system
      & Standard scaling
      & 1D, 2D \\
    \bottomrule
  \end{tabular}
  \caption{Input observables used in the different analysis configurations. All variables belong to the $t\bar{t}$ system. The `Transformation' column indicates the preprocessing applied before feeding the variable to the network.\label{tab:observables}}
\end{table}
%%%%%%%%%%%%%%%%%%%%%%%%%%%%%%%%%%%%%%%%%%%%%%
%
The signal hypothesis is characterised by one or more continuous parameters; in the presented case, these are $m_S$ and $C_e$. The signal dataset is generated discretely with the parameter values described in Tab.~\ref{tab:parameters}. These are concatenated with the observables to form the full input vector to the network, enabling the classifier to learn a parameter-dependent likelihood ratio $r(\mathbf{x} \mid \boldsymbol{\theta})$. 

The SM (class-0) events observables are also concatenated with the BSM parameters. Since the likelihood of SM events does not depend on $C_e$ or $m_S$, we apply a data augmentation technique to reflect this in the training. Each SM event is copied as many times as there are different signal parameters used for training. For example, when training the 1D configuration mass inference, we create 5 copies of each SM event, each copy concatenated with a different $m_S \in [600, 700, 800, 900, 1000]$.

%%%%%%%%%%%%%%%%%%%%%%%%%%%%%%%%%%%%%%%%%%%%%%
\begin{table}[!t]
  \centering
  \resizebox{\textwidth}{!}{
  \begin{tabular}{llll}
    \toprule
    \textbf{Parameter} & \textbf{Description} & \textbf{Training values} & \textbf{Holdout values} \\
    \midrule
    $C_e$
      & Signal coupling strength
      & 0.2, 0.4, 0.6, 0.8, 1.0, 1.2, 1.4, 1.6
      & 0.5, 1.1 \\
    $m_S$
      & Heavy Higgs boson mass [GeV]
      & 600, 700, 800, 900, 1000
      & 650, 870 \\
    \bottomrule
  \end{tabular}}
  \caption{Theory parameters used to condition the NN. The Training values column lists the discrete grid points at which simulated signal samples are available. The Holdout values are reserved for evaluation and not used during training. All parameters are scaled to $[0, 1]$ via min--max normalisation.\label{tab:parameters}}
\end{table}
%%%%%%%%%%%%%%%%%%%%%%%%%%%%%%%%%%%%%%%%%%%%%%

Monte Carlo events are assigned weights $w_i$ that account for cross section, luminosity (we use $140.1~\text{fb}^{-1}$ in this study), and generator-level effects. These can also be extended to the modelling of systematic uncertainties, for example, those due to detector effects not considered here. These weights can be negative due to interference between signal and background amplitudes, as explained in Sec.~\ref{sec:intf}. In actual experimental data, we can fit the background distribution by defining sidebands, extrapolating into the signal region whilst accounting for uncertainties, thereby yielding the reference class. Before training, the absolute event weights are normalised separately for each parameter-point category $C$ and weight sign:
\begin{equation}
  \label{eq:3.7}
  \tilde{w}_i^{(C, \pm)} \;=\;
  \frac{w_i}{\displaystyle\sum_{j \in \mathcal{S}_{C}^{\pm}} |w_j|}\,,
\end{equation}
where $\mathcal{S}_{C}^{+}$ ($\mathcal{S}_{C}^{-}$) denotes the subset of events in category $C$ with positive (negative) MC weights. This normalisation ensures that each parameter point contributes equally to the total ML model loss, preventing categories with larger cross sections from dominating the training.
The likelihood-ratio estimator is a fully connected feed-forward NN (a multilayer perceptron, MLP) with four layers, referred to as \texttt{RatioEstimatorNet}. The architecture is summarised in Tab.~\ref{tab:architecture}.

%%%%%%%%%%%%%%%%%%%%%%%%%%%%%%%%%%%%%%%%%%%%%%
\begin{table}[!b]
  \centering
  \begin{tabular}{clccc}
    \toprule
    \textbf{Layer} & \textbf{Type} & \textbf{Input dim.} & \textbf{Output dim.} & \textbf{Activation} \\
    \midrule
    1 & Fully connected & $d_{\mathbf{x}} + d_{\boldsymbol{\theta}}$ & $h$ & ReLU \\
    2 & Fully connected & $h$ & $h$ & ReLU \\
      & Dropout         & $h$ & $h$ & --- \\
    3 & Fully connected & $h$ & $h/2$ & ReLU \\
      & Dropout         & $h/2$ & $h/2$ & --- \\
    4 & Fully connected & $h/2$ & 1 & --- (logit) \\
    \bottomrule
  \end{tabular}
  \caption{Layer-by-layer description of the \texttt{RatioEstimatorNet} architecture. The input dimension is $d_{\mathbf{x}} + d_{\boldsymbol{\theta}}$, where $d_{\mathbf{x}}$ is the number of observables and $d_{\boldsymbol{\theta}}$ is the number of theory parameters. The hidden dimension $h$ is 256.\label{tab:architecture}}
\end{table}
%%%%%%%%%%%%%%%%%%%%%%%%%%%%%%%%%%%%%%%%%%%%%%
\newcommand{\Ne}{{N}_{\text{ev}}}
The network is trained as a binary classifier using the weighted binary cross-entropy (BCE) loss with logits. For a mini-batch of $\Ne$ events, the loss is:
\begin{equation}
  \mathcal{L}
  \;=\;
  \sum_{i=1}^{\Ne} \bigl|\tilde{w}_i\bigr|
  \bigl[
    -y_i \log\sigma(f_i)
    - (1 - y_i) \log\bigl(1 - \sigma(f_i)\bigr)
  \bigr]\,,
  \label{eq:loss}
\end{equation}
where $f_i = f(\mathbf{x}_i, \boldsymbol{\theta}_i)$ is the network output logit, $y_i \in \{0, 1\}$ is the class label, $\sigma(\cdot)$ is the sigmoid function, and $\tilde{w}_i$ are the normalised luminosity weights obtained from Eq.~\eqref{eq:3.7}. We use {\tt{PyTorch}}~\cite{Paszke:2019xhz} for training, {\tt{scikit-learn}}~\cite{Pedregosa:2011ork} for data processing, and {\tt{PyTorch Lightning}}~\cite{pytorchl} to standardise and structure the deep learning code. The network hyperparameters for training are summarised in Tab.~\ref{tab:hyperparameters}.

%%%%%%%%%%%%%%%%%%%%%%%%%%%%%%%%%%%%%%%%%%%%%%
\begin{table}[!t]
  \centering
  \begin{tabular}{lccccccc}
    \toprule
    \textbf{Configuration}
      & \textbf{Input dim.}
      & $h$
      & \textbf{Epochs}
      & \textbf{Batch size}
      & \textbf{LR}
      & \textbf{LR patience}
      & \textbf{LR factor} \\
    \midrule
    1D coupling  & $4 + 1 = 5$ & 256  & 150 & 10\,240 & $2 \times 10^{-4}$ & 10 & 0.8 \\
    1D mass      & $4 + 1 = 5$ & 256  & 160 & 10\,240 & $2 \times 10^{-4}$ & 10 & 0.8 \\
    2D           & $4 + 2 = 6$ & 256  & 300 & 20\,480 & $4 \times 10^{-4}$ & 10 & 0.8 \\
    \bottomrule
  \end{tabular}
  \caption{Hyperparameters for the different analysis configurations. All configurations use the {\tt{AdamW}} optimiser, a dropout rate of $p = 0.05$, weight decay of $10^{-3}$, and the \texttt{ReduceLROnPlateau} scheduler. The `Input dim.' column shows $d_{\mathbf{x}} + d_{\boldsymbol{\theta}}$.\label{tab:hyperparameters}}
\end{table}
%%%%%%%%%%%%%%%%%%%%%%%%%%%%%%%%%%%%%%%%%%%%%%

% --- TikZ figure: RoSMM Inference Pipeline ---
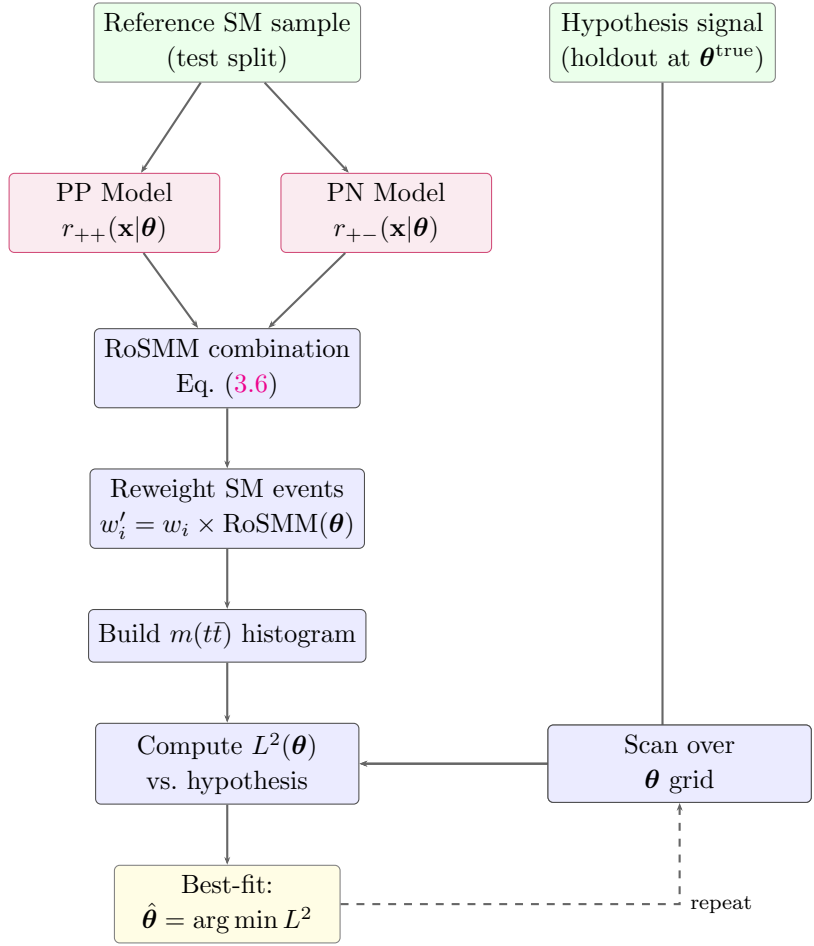
\begin{figure}[htbp]
  \centering
  \begin{tikzpicture}[
      node distance=0.9cm and 2.5cm,
      every node/.style={font=\small},
      block/.style={
        rectangle, draw=black!70, fill=blue!8,
        minimum width=3.5cm, minimum height=0.7cm,
        rounded corners=2pt, align=center
      },
      model/.style={
        rectangle, draw=purple!70, fill=purple!8,
        minimum width=2.8cm, minimum height=0.7cm,
        rounded corners=2pt, align=center
      },
      data/.style={
        rectangle, draw=black!50, fill=green!8,
        minimum width=3.0cm, minimum height=0.7cm,
        rounded corners=2pt, align=center
      },
      result/.style={
        rectangle, draw=black!50, fill=yellow!12,
        minimum width=3.0cm, minimum height=0.7cm,
        rounded corners=2pt, align=center
      },
      arrow/.style={-{Stealth[length=3pt]}, thick, draw=black!60},
    ]

    % Data sources
    \node[data] (ref) {Reference SM sample\\(test split)};
    \node[data, right=2.5cm of ref] (hyp) {Hypothesis signal\\(holdout at $\boldsymbol{\theta}^{\mathrm{true}}$)};

    % Models
    \node[model, below=1.2cm of ref, xshift=-1.5cm] (pp) {PP Model\\$r_{++}(\mathbf{x}\mid\boldsymbol{\theta})$};
    \node[model, right=0.8cm of pp] (pn) {PN Model\\$r_{+-}(\mathbf{x}\mid\boldsymbol{\theta})$};

    % RoSMM combination
    \node[block, below=1.0cm of pp, xshift=1.5cm] (rosmm) {RoSMM combination\\Eq.~(\ref{eq:3.6})};

    % Reweighting
    \node[block, below=0.8cm of rosmm] (rew) {Reweight SM events\\$w_i' = w_i \times \mathrm{RoSMM}(\boldsymbol{\theta})$};

    % Histogram
    \node[block, below=0.8cm of rew] (hist) {Build $\mttbar$ histogram};

    % Chi2
    \node[block, below=0.8cm of hist] (chi2) {Compute $L^2(\boldsymbol{\theta})$\\vs.\ hypothesis};

    % Scan loop
    \node[block, right=2.5cm of chi2] (scan) {Scan over\\$\boldsymbol{\theta}$ grid};

    % Result
    \node[result, below=0.8cm of chi2] (bestfit) {Best-fit:\\$\hat{\boldsymbol{\theta}} = \arg\min L^2$};

    % Arrows
    \draw[arrow] (ref) -- (pp);
    \draw[arrow] (ref) -- (pn);
    \draw[arrow] (pp) -- (rosmm);
    \draw[arrow] (pn) -- (rosmm);
    \draw[arrow] (rosmm) -- (rew);
    \draw[arrow] (rew) -- (hist);
    \draw[arrow] (hist) -- (chi2);
    \draw[arrow] (hyp) |- (chi2);
    \draw[arrow] (chi2) -- (bestfit);
    \draw[arrow] (scan.west) -- (chi2.east);
    \draw[arrow, dashed] (bestfit.east) -| (scan.south) node[midway, right, font=\scriptsize] {repeat};
    \node[block, right=2.5cm of chi2] (scan2) {Scan over\\$\boldsymbol{\theta}$ grid};

  \end{tikzpicture}
  \caption{Schematic of the RoSMM inference pipeline. Two pre-trained models (PP and PN) are combined to reweigh a reference background sample at each scan point $\boldsymbol{\theta}$. The reweighted $\mttbar$ distribution is compared to the holdout signal hypothesis via an $L^2$ statistic. The best-fit parameter is determined by minimising $L^2$ over the scan grid.\label{fig:rosmm_pipeline}}
\end{figure}

%%%%%%%%%%%%%%%%%%%%%%%%%%%%%%%%%%%%%%%%%%%%%%
\subsection{Inference}
%%%%%%%%%%%%%%%%%%%%%%%%%%%%%%%%%%%%%%%%%%%%%%
For parameter inference, the two independently trained sub-density ratio models, the \textit{PP model} and the PN model, are combined according to Eq.~\eqref{eq:3.6}. Inference then proceeds as follows:
\begin{enumerate}
  \item \textbf{Reference sample}: A set of SM events is
    selected from the test split of the dataset.
  \item \textbf{Hypothesis construction}: The holdout signal events at
    the true parameter point $\boldsymbol{\theta}^{\mathrm{true}}$ are
    histogrammed in the observable of interest, $m_{t\bar{t}}$ for this work,
    to form the hypothesis distribution~$H$.
  \item \textbf{Parameter scan}: For each point $\boldsymbol{\theta}$ on
    a scan grid:
    \begin{enumerate}
      \item The RoSMM reweighting factor
        (Eq.~\eqref{eq:3.6}) is computed for each reference event.
      \item The reference events are reweighted by
        $w_i' = w_i \times \mathrm{RoSMM}(\boldsymbol{\theta}
        \mid \mathbf{x}_i)$.
      \item The reweighted events are histogrammed to form the inference
        distribution $I(\boldsymbol{\theta})$.
      \item An $L^2$ statistic is computed as
        \begin{equation}
          L^2(\boldsymbol{\theta})
          \;=\;
          \sum_{b=1}^{N_{\mathrm{bins}}}
          \bigl(H_b - I_b(\boldsymbol{\theta})\bigr)^2\,.
          \label{eq:chi2}
        \end{equation}
    \end{enumerate}
  \item \textbf{Best fit}: The best-fit parameter is the one that
    minimises $L^2$:
    \begin{equation}
      \hat{\boldsymbol{\theta}}
      \;=\;
      \arg\min_{\boldsymbol{\theta}} \; L^2(\boldsymbol{\theta})\,.
      \label{eq:bestfit}
    \end{equation}
\end{enumerate}

%%%%%%%%%%%%%%%%%%%%%%%%%%%%%%%%%%%%%%%%%%%%%%
%\subsection{Uncertainty Estimation}
\label{sec:pseudo_experiments}
%%%%%%%%%%%%%%%%%%%%%%%%%%%%%%%%%%%%%%%%%%%%%%
We estimate statistical uncertainties on the best-fit parameters using a pseudo-experiment method. Starting from the hypothesis distribution $H$ with bin-by-bin statistical uncertainties $\sigma_b$, a set of $N_{\mathrm{PE}} = 1k$ pseudo-experiments is generated by independently fluctuating each bin according to a normal distribution
\begin{equation}
  H_b^{(k)} \text{~sampled from~} \mathcal{N}(H_b,\; \sigma_b)\ ,
  \quad k = 1, \ldots , N_{\mathrm{PE}}\ .
  \label{eq:pseudo}
\end{equation}
For each pseudo-experiment $k$, the best-fit parameter $\hat{\boldsymbol{\theta}}^{(k)}$ is determined by minimising the $L^2$ statistic, Eq.~\eqref{eq:chi2}, with $H$ replaced by $H^{(k)}$. The central 95\% confidence interval for each parameter is obtained from the distribution of $\hat{\boldsymbol{\theta}}^{(k)}$ values:
\begin{equation}
  \left[
    \hat{\theta}_{\alpha/2}^{(k)},\;
    \hat{\theta}_{1-\alpha/2}^{(k)}
  \right]\,,
  \qquad \alpha = 0.05\,.
  \label{eq:confidence_interval}
\end{equation}

%%%%%%%%%%%%%%%%%%%%%%%%%%%%%%%%%%%%%%%%%%%%%%
\section{Results}
\label{sec:res}
%%%%%%%%%%%%%%%%%%%%%%%%%%%%%%%%%%%%%%%%%%%%%%
With the inference pipeline in place, we analyse the performance of our approach in three successive steps. We first infer subsets of information (the coupling strength $C_e$ or the mass $m_S$ in the new physics scenario). These 1D inference configurations serve as validation of the method but also as a testing ground to understand the limitations of our methodology. Secondly, we  extract the model parameters simultaneously and, in a third step, comment on the algorithm's robustness to changes in our model's correlations. The latter is achieved by changing the resonance width away from the prediction of Eq.~\eqref{eq:4.1}; we will consider perturbative benchmarks of $\Gamma_S/m_S$ as in typical experimental analyses, see for example the ATLAS search of Ref.~\cite{ATLAS:2025kmo}. 

%%%%%%%%%%%%%%%%%%%%%%%%%%%%%%%%%%%%%%%%%%%%%%
\begin{figure}[!b]
    \centering
    \subfigure[]{\includegraphics[width=.48\linewidth]{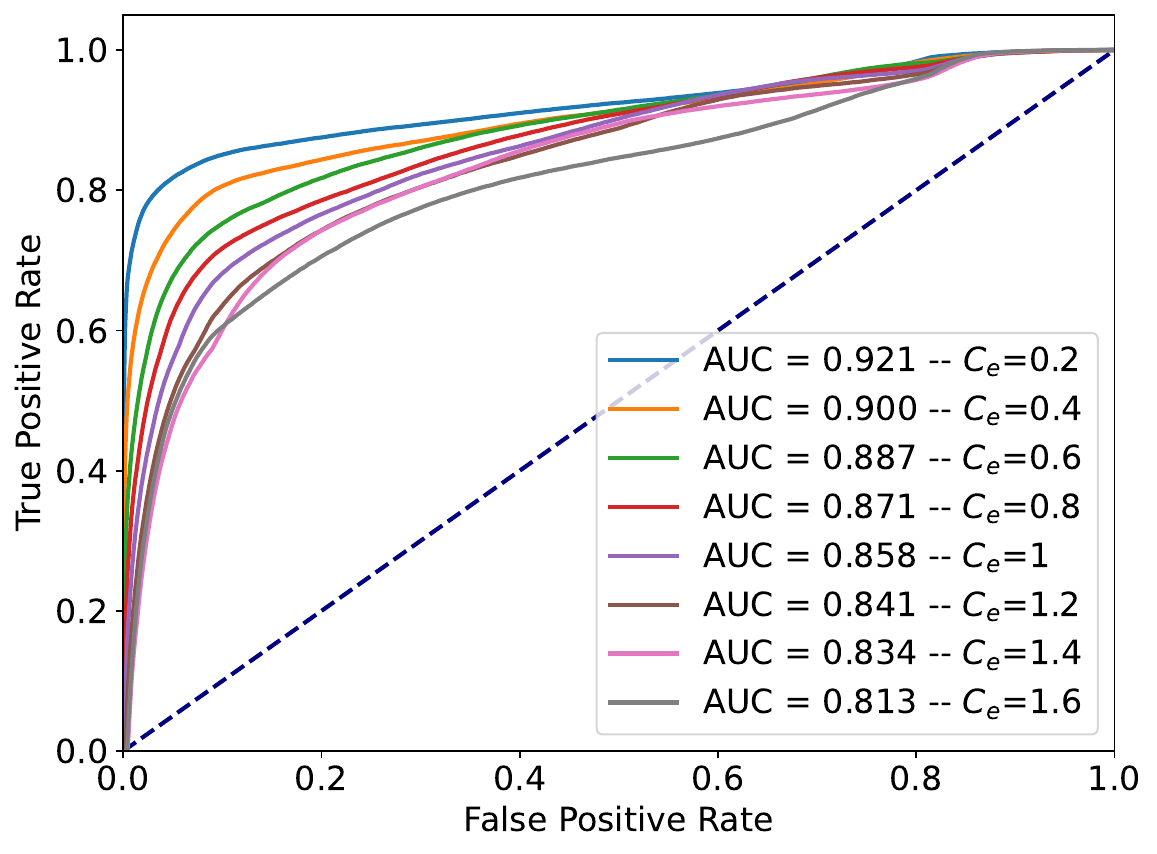}}
    \hfill
    \subfigure[]{\includegraphics[width=.48\linewidth]{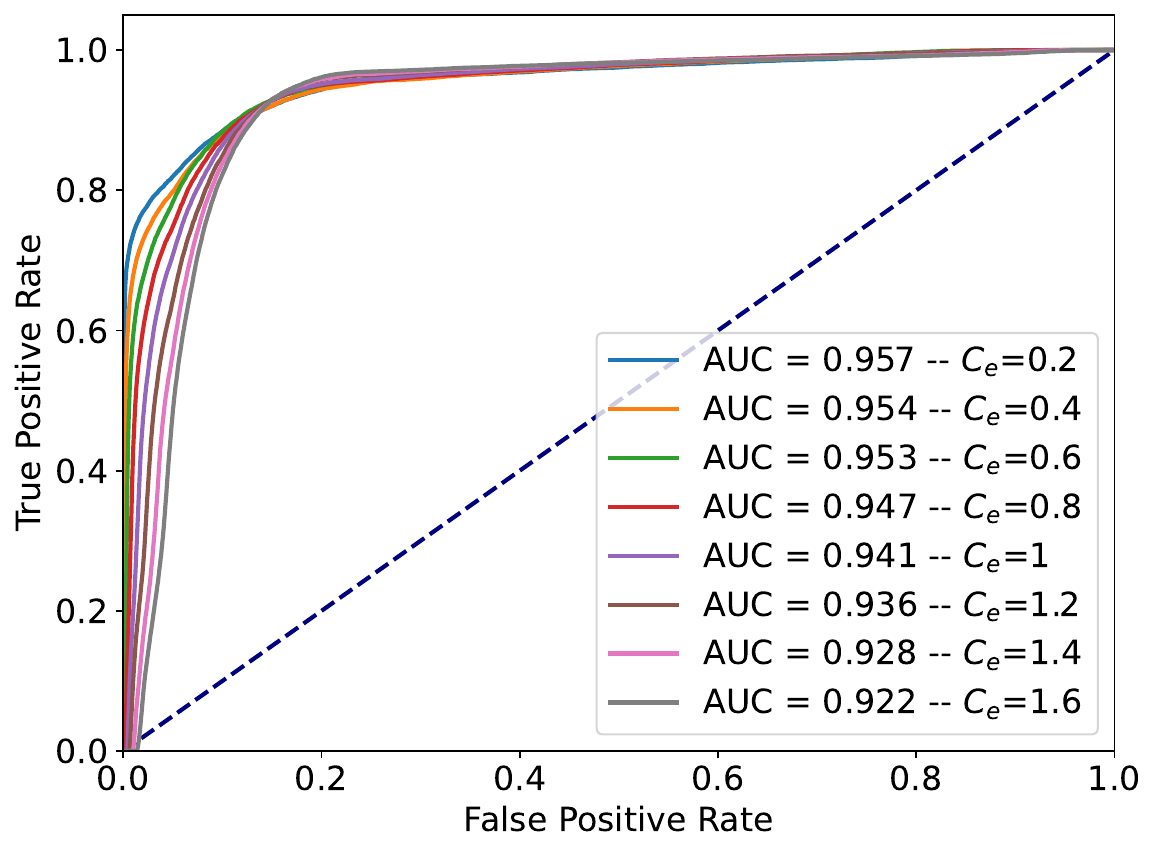}}
    \caption{ROC curves for each sub-density ratio model: (a) $r_{++}$ and (b) $r_{+-}$. Each $C_e$ hypothesis is shown separately in the plots, including the area-under-curve (AUC) values, demonstrating excellent separation.\label{fig:performance_1D_coupling}}
\end{figure}
%%%%%%%%%%%%%%%%%%%%%%%%%%%%%%%%%%%%%%%%%%%%%%
%%%%%%%%%%%%%%%%%%%%%%%%%%%%%%%%%%%%%%%%%%%%%%
\begin{figure}[!t]
    \centering
    \subfigure[]{\includegraphics[width=.48\linewidth]{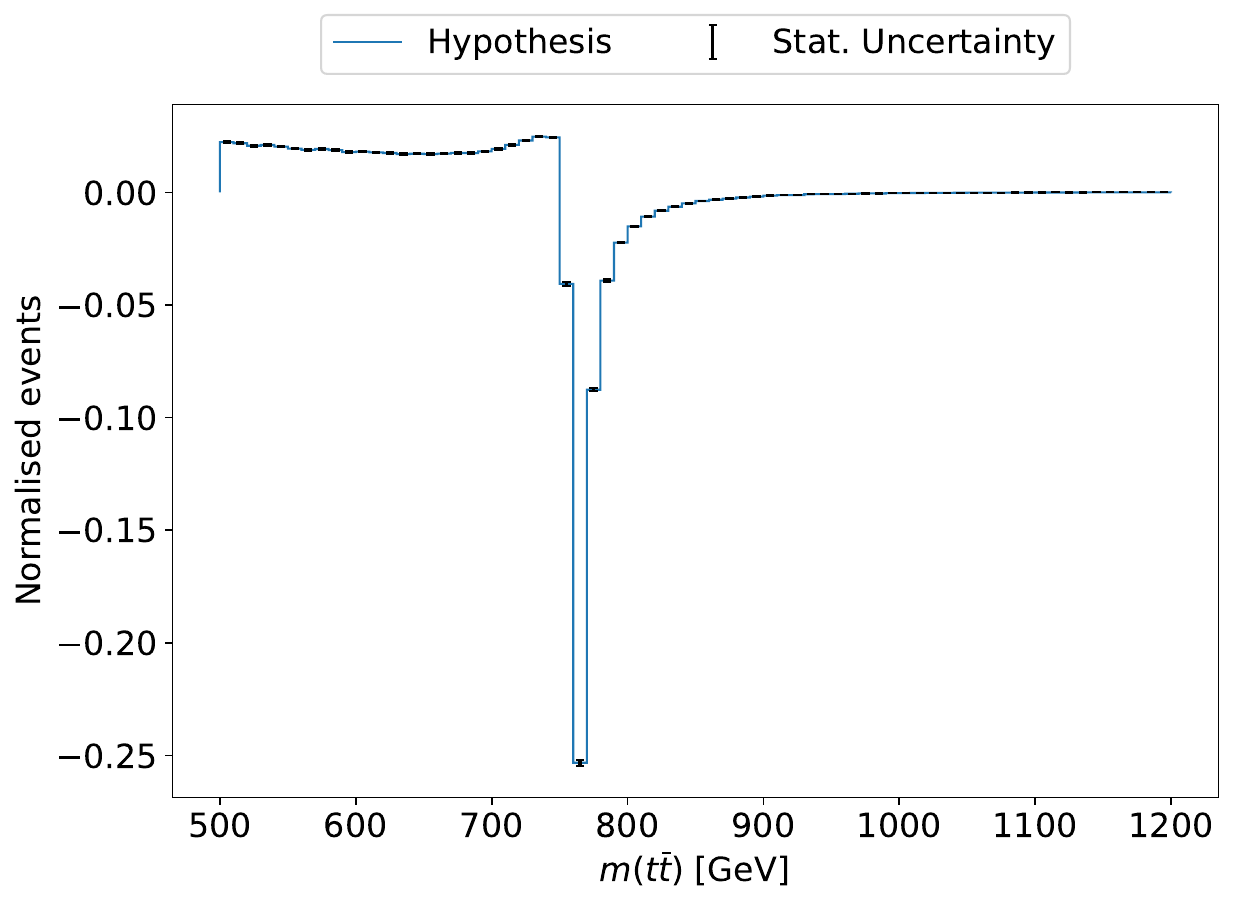}}
    \hfill
    \subfigure[]{\includegraphics[width=.48\linewidth]{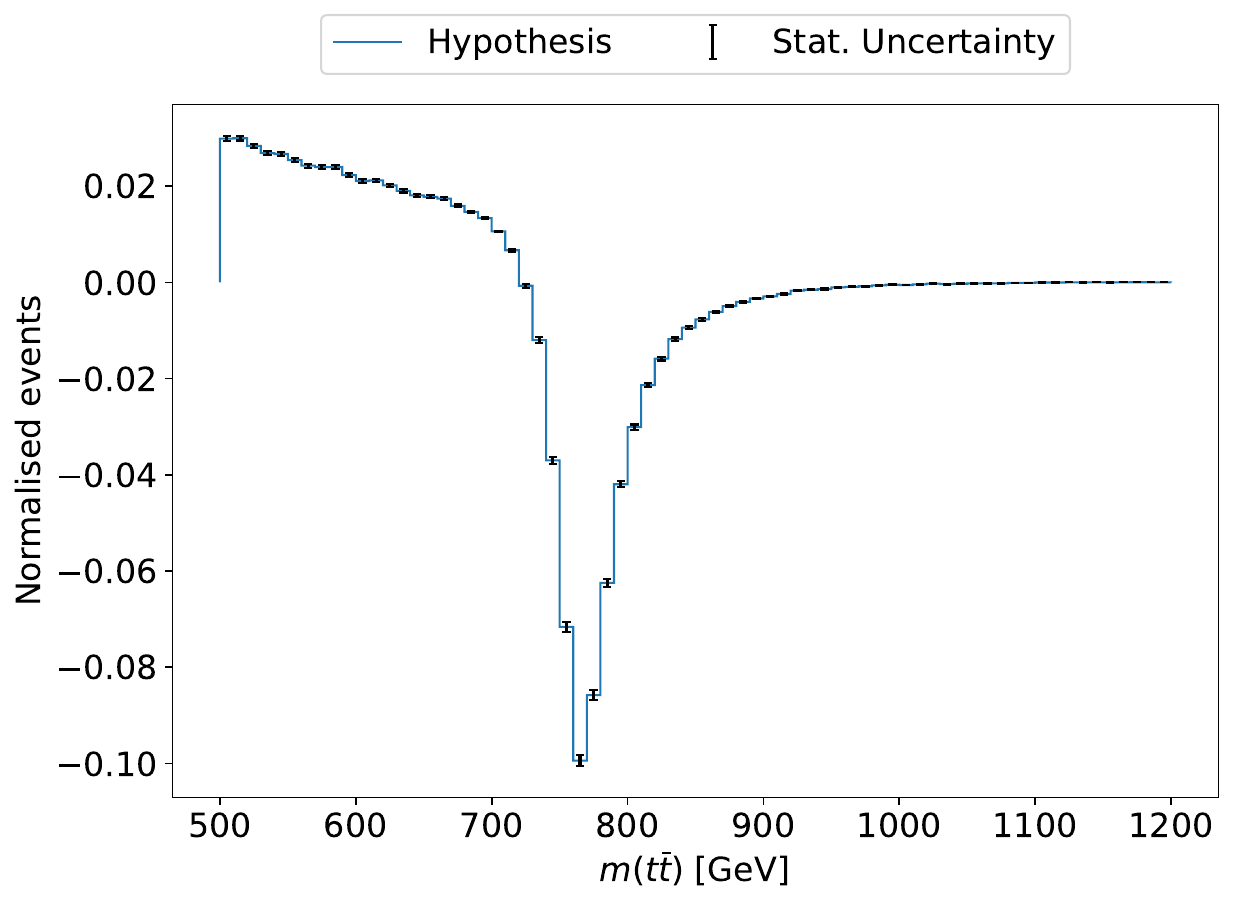}}
    \caption{Distribution of $\mttbar$ for (a) $C_e=0.5$ and (b) $C_e=1.1$. The plots show the distribution densities, with the histograms' absolute areas normalised to unity.\label{fig:hypothesis_1D_coupling}}
\end{figure}
%%%%%%%%%%%%%%%%%%%%%%%%%%%%%%%%%%%%%%%%%%%%%%
%%%%%%%%%%%%%%%%%%%%%%%%%%%%%%%%%%%%%%%%%%%%%%
\begin{figure}[!b]
    \centering
    \subfigure[]{\includegraphics[width=.48\linewidth]{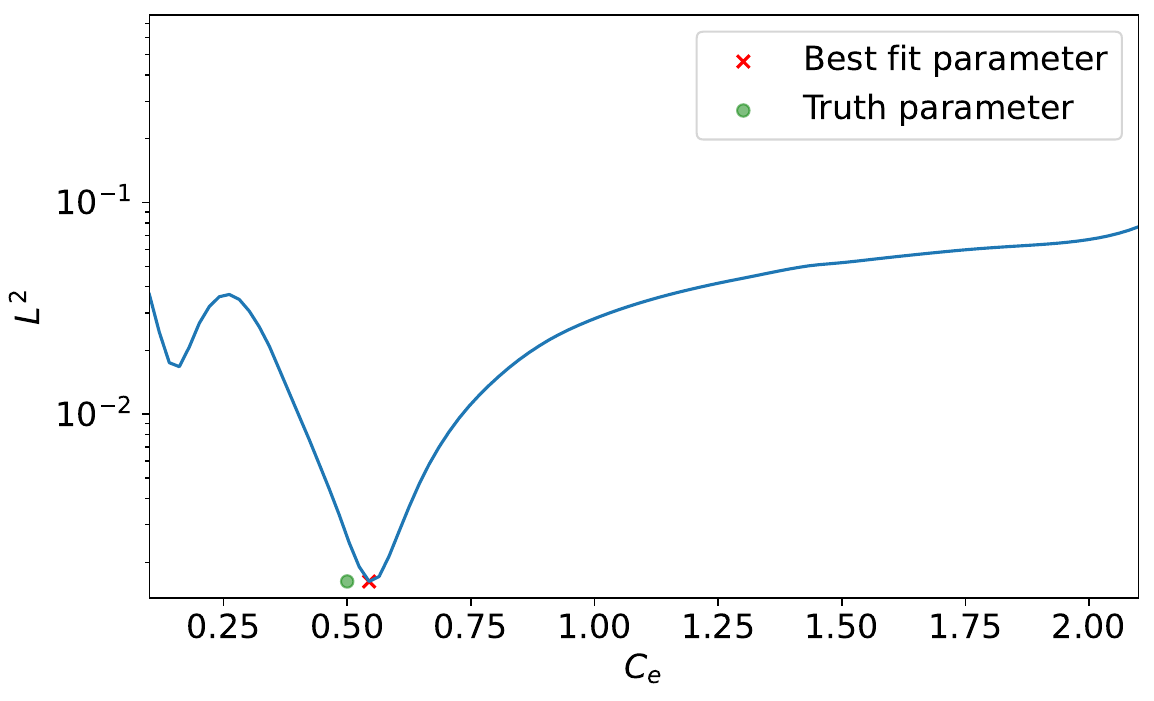}}
    \hfill
    \subfigure[]{\includegraphics[width=.48\linewidth]{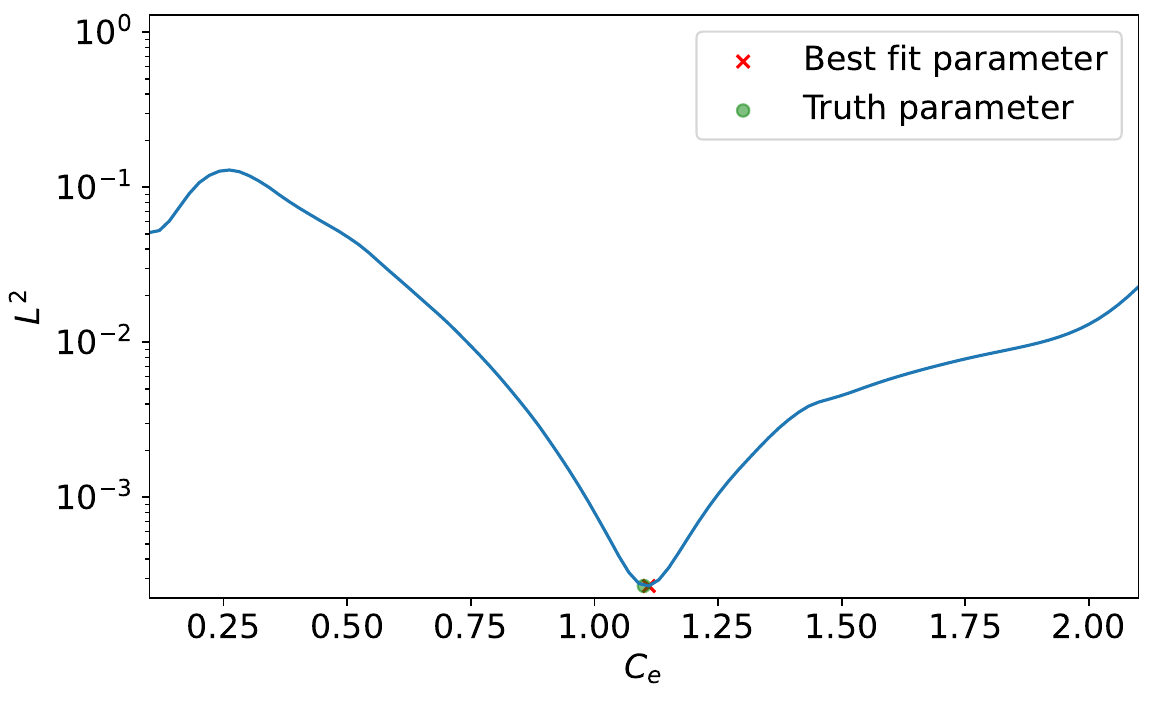}}
    \caption{$L^2$ scans for (a) $C_e=0.5$ and (b) $C_e=1.1$ hypothesis. The green circle shows the truth $C_e$ value, and the red cross shows the estimated value via the inference pipeline.\label{fig:scan_1D_coupling}}
\end{figure}
%%%%%%%%%%%%%%%%%%%%%%%%%%%%%%%%%%%%%%%%%%%%%%
%%%%%%%%%%%%%%%%%%%%%%%%%%%%%%%%%%%%%%%%%%%%%%
\begin{figure}[!t]
    \centering
    \subfigure[]{\includegraphics[width=.48\linewidth]{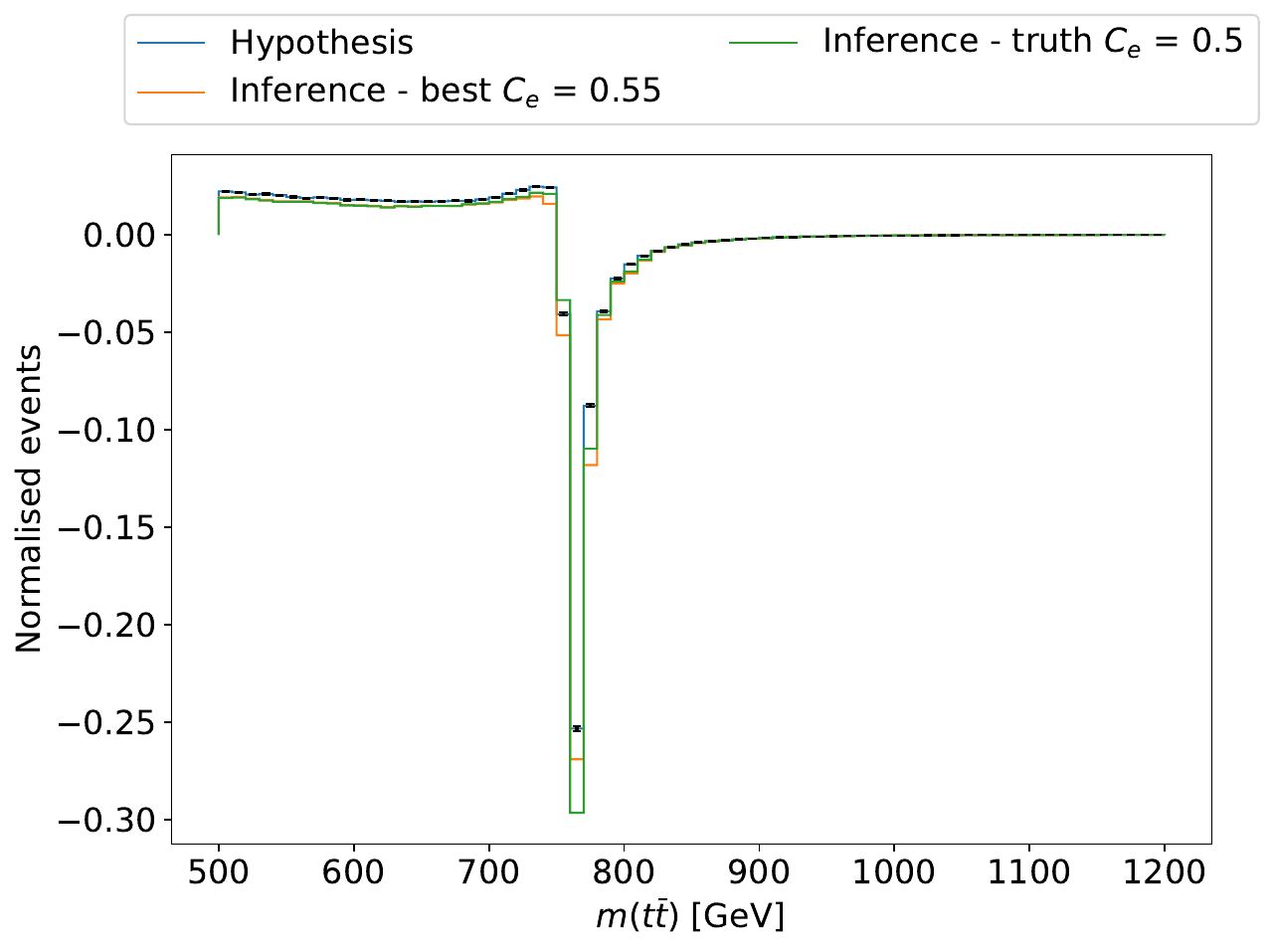}}
    \hfill
    \subfigure[]{\includegraphics[width=.48\linewidth]{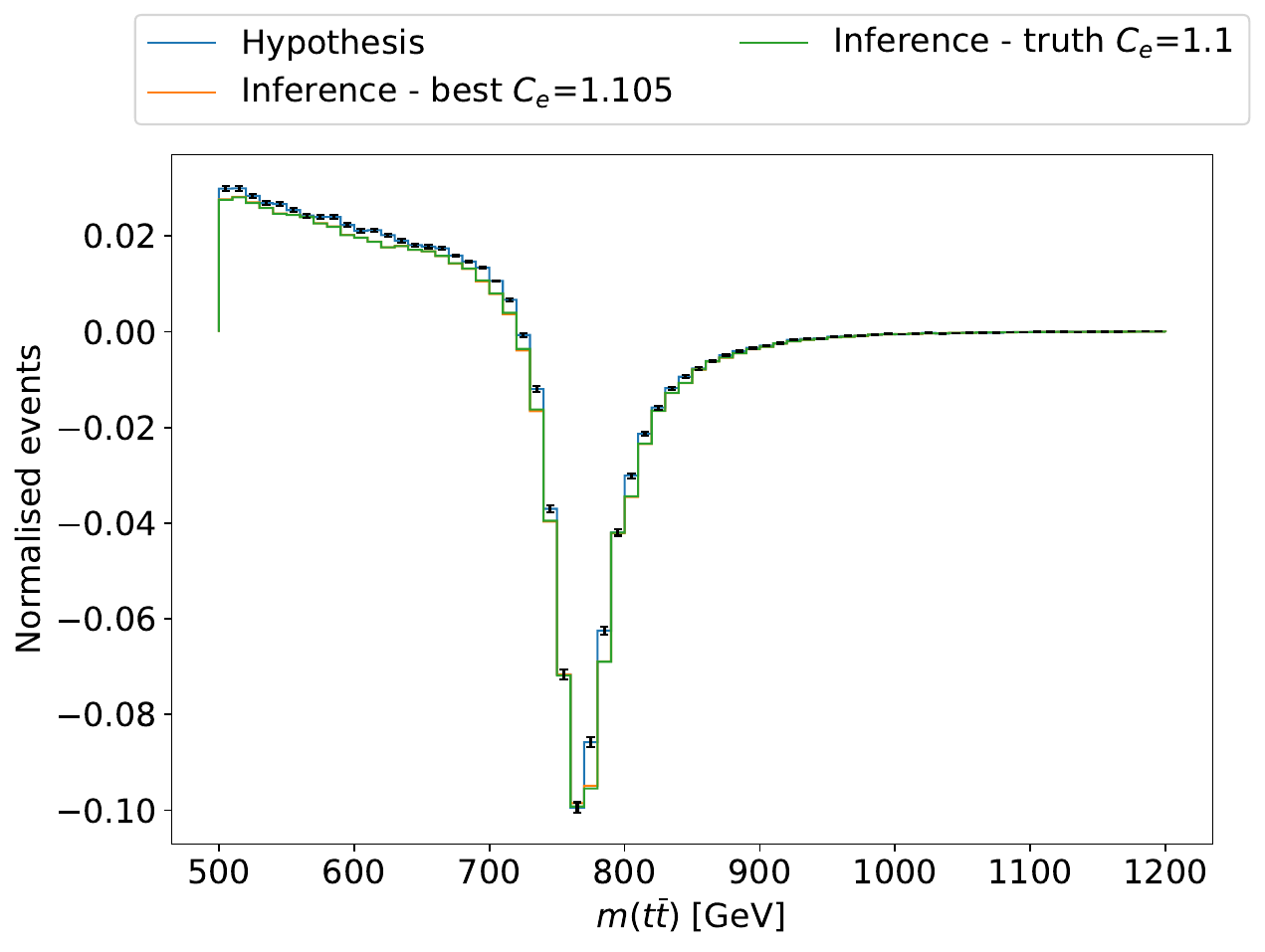}}
    \caption{Inference pipeline summary for (a) $C_e=0.5$ and (b) $C_e=1.1$. In each plot, the blue line shows the truth hypothesis $\mttbar$ distribution, the orange line shows the best-fit parameter estimated via the inference pipeline, and the green line shows the distribution generated from performing inference with the truth value for the $C_e$ parameter. All the distributions are normalised to unity.\label{fig:summary_1D_coupling}}
\end{figure}
%%%%%%%%%%%%%%%%%%%%%%%%%%%%%%%%%%%%%%%%%%%%%%
%%%%%%%%%%%%%%%%%%%%%%%%%%%%%%%%%%%%%%%%%%%%%%
\begin{figure}[!b]
    \centering
    \subfigure[]{\includegraphics[width=.48\linewidth]{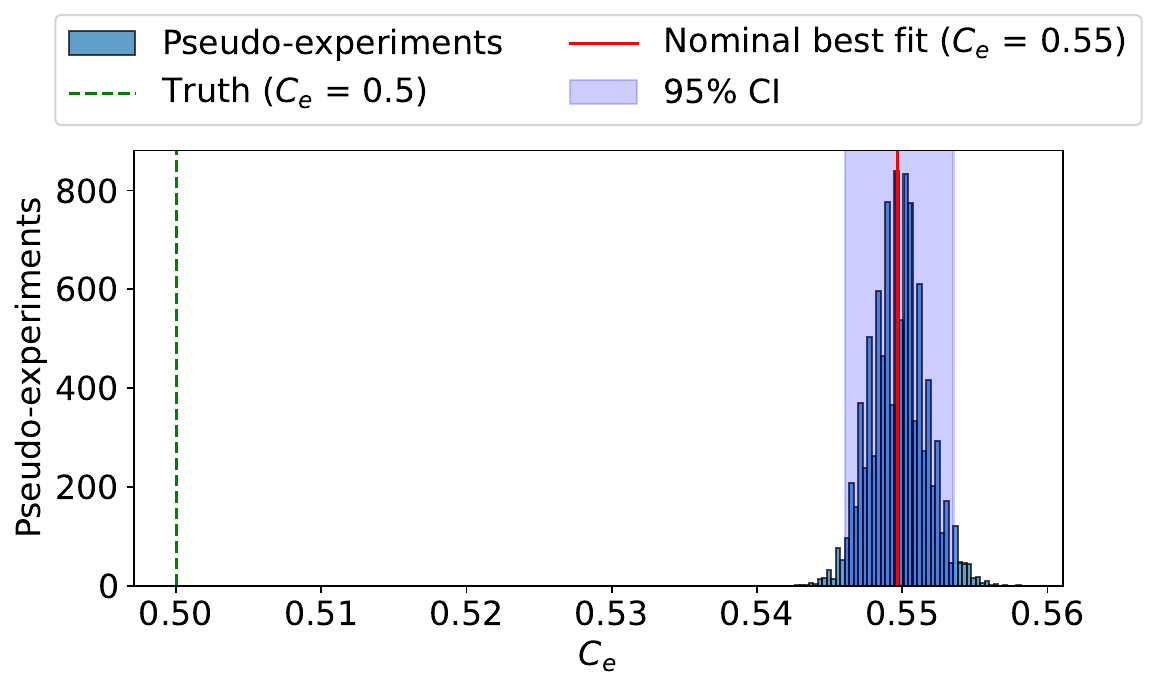}}
    \hfill
    \subfigure[]{\includegraphics[width=.48\linewidth]{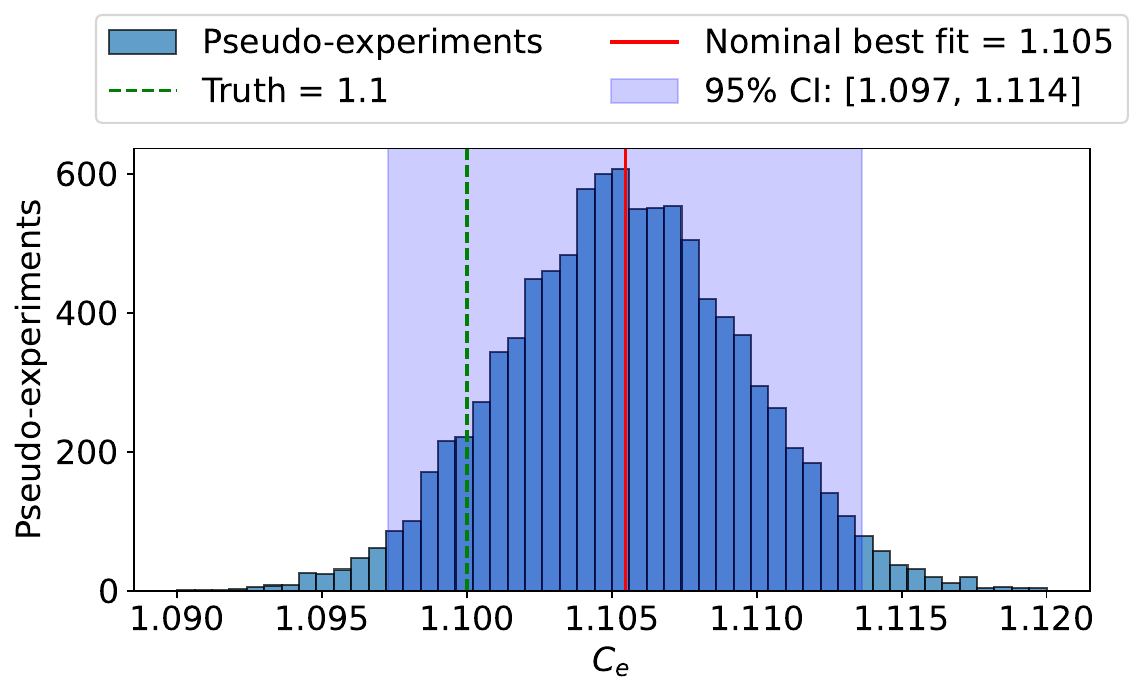}}
    \caption{Pseudo-experiments results for (a) $C_e=0.5$ and (b) $C_e=1.1$. The green dotted line shows the truth $C_e$ value, the red vertical line shows the estimated parameter value, and the blue band shows where 95\% of the pseudo-experiments are concentrated.}
    \label{fig:uncertainty_1D_coupling}
\end{figure}
%%%%%%%%%%%%%%%%%%%%%%%%%%%%%%%%%%%%%%%%%%%%%%
%%%%%%%%%%%%%%%%%%%%%%%%%%%%%%%%%%%%%%%%%%%%%%
\begin{figure}[!t]
    \centering
    \subfigure[]{\includegraphics[width=.68\linewidth]{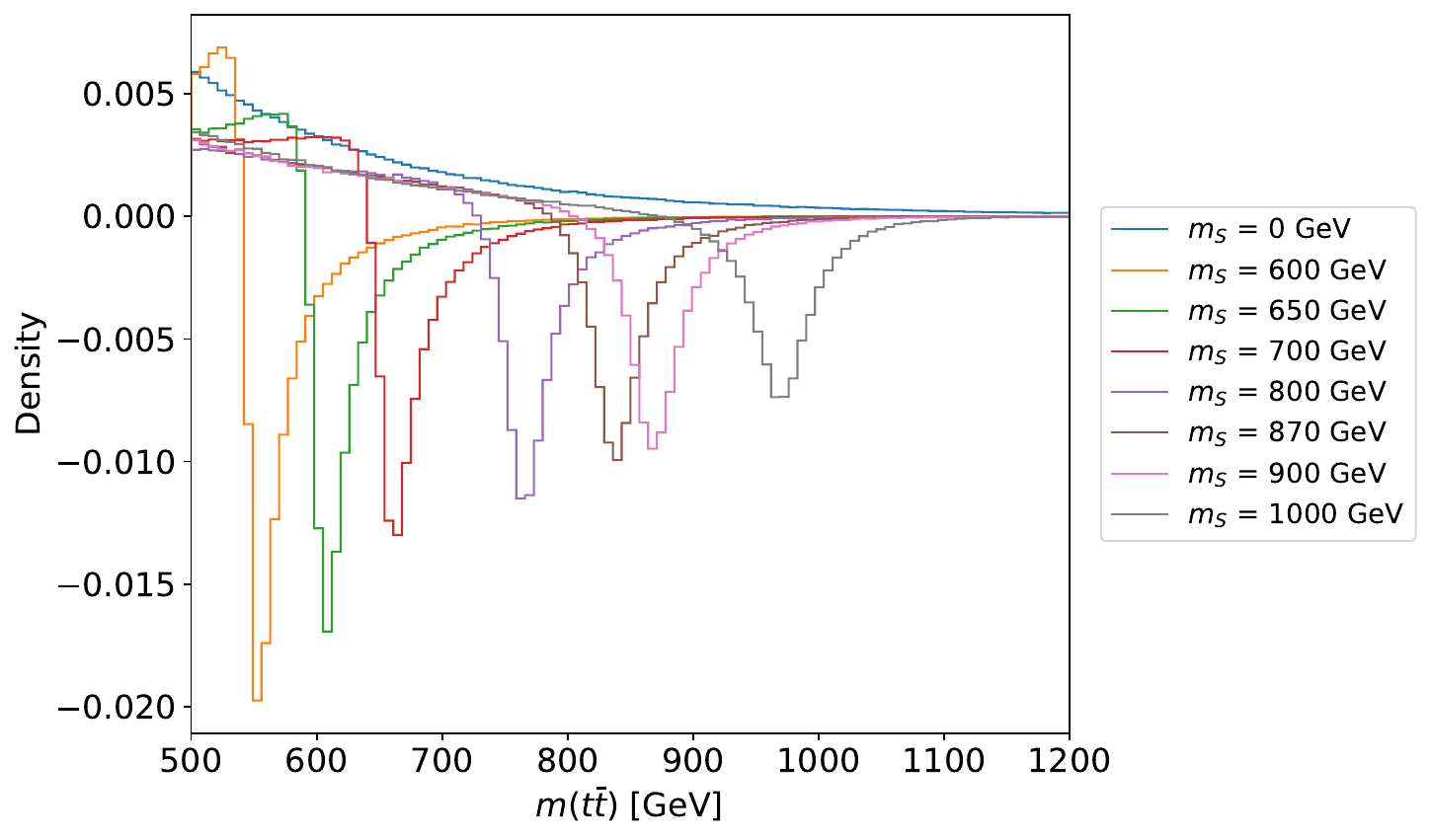}}
    \caption{Distribution of $\mttbar$ for all $m_S$ values used in training and inference split by hypothesis. The $m_S=0$ shows the SM distribution.\label{fig:mttbar_1D_mass_all}}
\end{figure}
%%%%%%%%%%%%%%%%%%%%%%%%%%%%%%%%%%%%%%%%%%%%%%

%%%%%%%%%%%%%%%%%%%%%%%%%%%%%%%%%%%%%%%%%%%%%%
\subsection{Testing the framework: 1D inference configurations}
\label{sec:res1d}
%%%%%%%%%%%%%%%%%%%%%%%%%%%%%%%%%%%%%%%%%%%%%%
We train two different pipelines; each performs parameter estimation under a single-parameter signal hypothesis. 
%%%%%%%%%%%%%%%%%%%%%%%%%%%%%%%%%%%%%%%%%%%%%%
\subsubsection{Inference of $C_e$}
\label{sec:ce}
%%%%%%%%%%%%%%%%%%%%%%%%%%%%%%%%%%%%%%%%%%%%%%
We train our two sub-density ratio estimator models according to the architecture described in Sec.~\ref{sec:arch}. In this part, we fix $m_S$ to 800 GeV. The dataset consists of 830k class-0 events, and 
between 746k and 823k class-1 events for $C_e = 1.6$ and $C_e = 0.4$, depending on the MC efficiency. The total number of class-1 events is approximately 7.8m. The coupling values used for training and inference validation (holdout) are shown in Tab.~\ref{tab:parameters}. For this configuration, all the signal hypotheses were generated with $m_S = 800$ GeV.

Once each sub-density ratio estimator is trained, we calculate the receiver operating characteristic (ROC) curves shown in Fig.~\ref{fig:performance_1D_coupling}, with each signal hypothesis displayed separately. The models achieve better performance at lower coupling values. In this region, the signal contribution becomes more concentrated in $\mttbar$, which aids the discrimination.

We use the RoSMM model and the pipeline outlined in Fig.~\ref{fig:rosmm_pipeline} to infer the coupling value assuming hypothetical data from each of the holdout coupling hypotheses shown in Tab.~\ref{tab:parameters} ($C_e = 0.5, 1.1$). These two datasets are never seen by the models for the RoSMM training. The $\mttbar$ shapes for these two samples are shown in Fig.~\ref{fig:hypothesis_1D_coupling}. The shapes of the $L^2$ scans for $C_e \in [0.1,2.1]$ can be seen in Fig.~\ref{fig:scan_1D_coupling}, with the truth value marked by the green circle and the inference result with the red cross. We observe a clear minimum in the $L^2$ statistic. Figure~\ref{fig:summary_1D_coupling} presents the comparison between shapes of the hypothesis, the shape given by our RoSMM model at the estimated $C_e$, and the shape given by the RoSMM model at the truth $C_e$. Finally, to estimate the statistical uncertainty of our procedure, we perform 10k pseudo-experiments as described in Sec.~\ref{sec:pseudo_experiments}, and repeat the $L^2$ scans to obtain the uncertainty band. The results of this procedure can be seen in Fig.~\ref{fig:uncertainty_1D_coupling}. We will comment on our statistical observations in Sec.~\ref{sec:ms}, directly comparing with the extraction of $m_S$ (at fixed $C_e$).

%%%%%%%%%%%%%%%%%%%%%%%%%%%%%%%%%%%%%%%%%%%%%%
\subsubsection{Inference of $m_S$}
\label{sec:ms}
%%%%%%%%%%%%%%%%%%%%%%%%%%%%%%%%%%%%%%%%%%%%%%
To infer the resonance mass in isolation (at fixed $C_e$), the dataset is formed of 800k class-0 events, and 736k to 798k class-1 events for hypothesis $m_S = 1000~\text{GeV}$ and $m_S = 700~\text{GeV}$, depending on MC efficiency. (The total number of class-1 events is 5.4m.) The mass values used for training and inference validation (holdout values) are shown in Tab.~\ref{tab:parameters}. For this configuration, all the signal hypotheses were generated with $C_e = 1.0$.

Examples of the $\mttbar$ distribution for different $m_S$ hypotheses are shown in Fig.~\ref{fig:mttbar_1D_mass_all}. Once each sub-density ratio estimator is trained, we calculate the ROC curves, yielding very good discrimination with AUC values around $0.85-0.9$ depending on the mass hypothesis.We observe that the $r_{++}$ model performs better at lower $m_S$ values. This is because the positive signal contributions become more concentrated at lower $\mttbar$ values, as can be seen in Fig.~\ref{fig:mttbar_1D_mass_all}. For the $r_{+-}$ model, since the depth of the interference dip concentrates as $m_S$ increases, we observe better performance at higher $m_S$.

%%%%%%%%%%%%%%%%%%%%%%%%%%%%%%%%%%%%%%%%%%%%%%
\begin{figure}[!t]
    \centering
    \subfigure[]{\includegraphics[width=.48\linewidth]{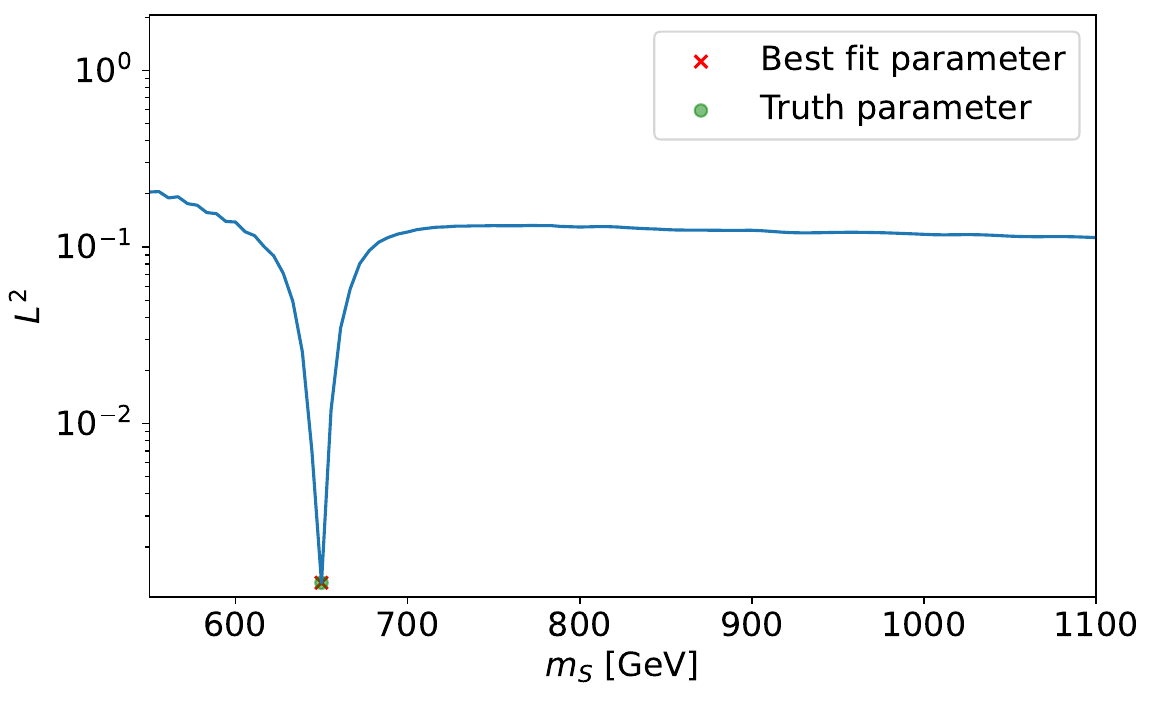}}
    \hfill
    \subfigure[]{\includegraphics[width=.48\linewidth]{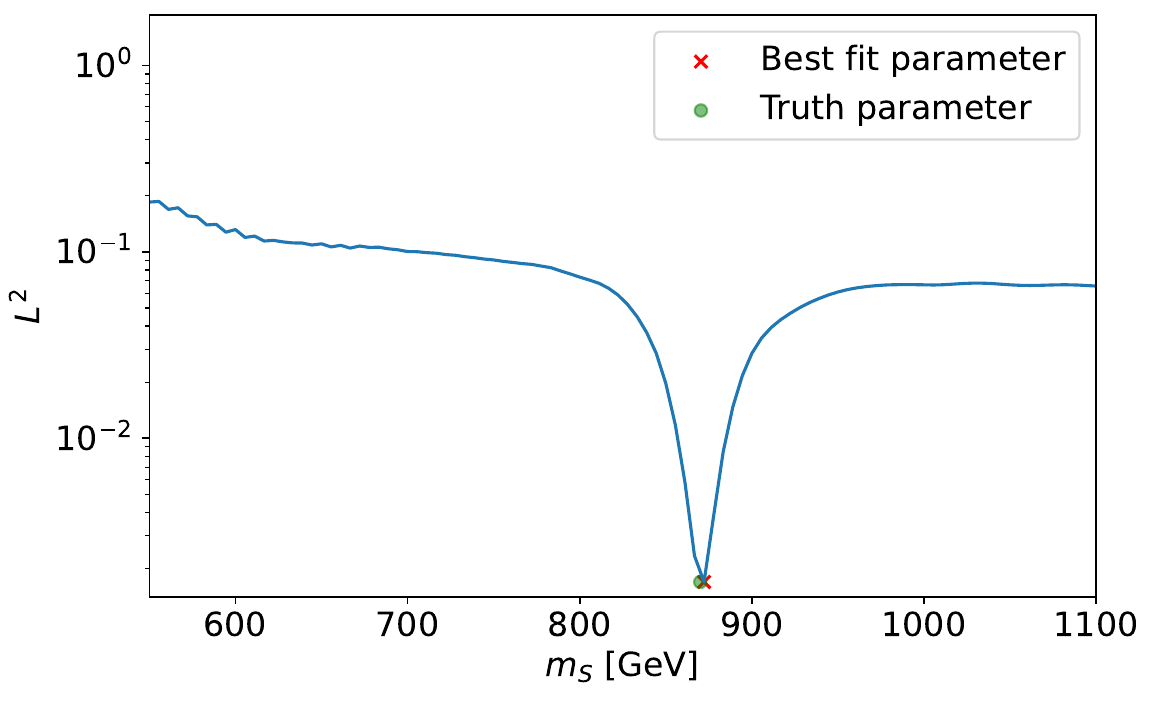}}
    \caption{$L^2$ scans for (a) $650$ and (b) $870$ GeV hypothesis. The green circle shows the truth $m_S$ value, and the red cross shows the estimated value via the inference pipeline.\label{fig:scan_1D_mass}}
\end{figure}
\begin{figure}[!b]
    \centering
    \subfigure[]{\includegraphics[width=.48\linewidth]{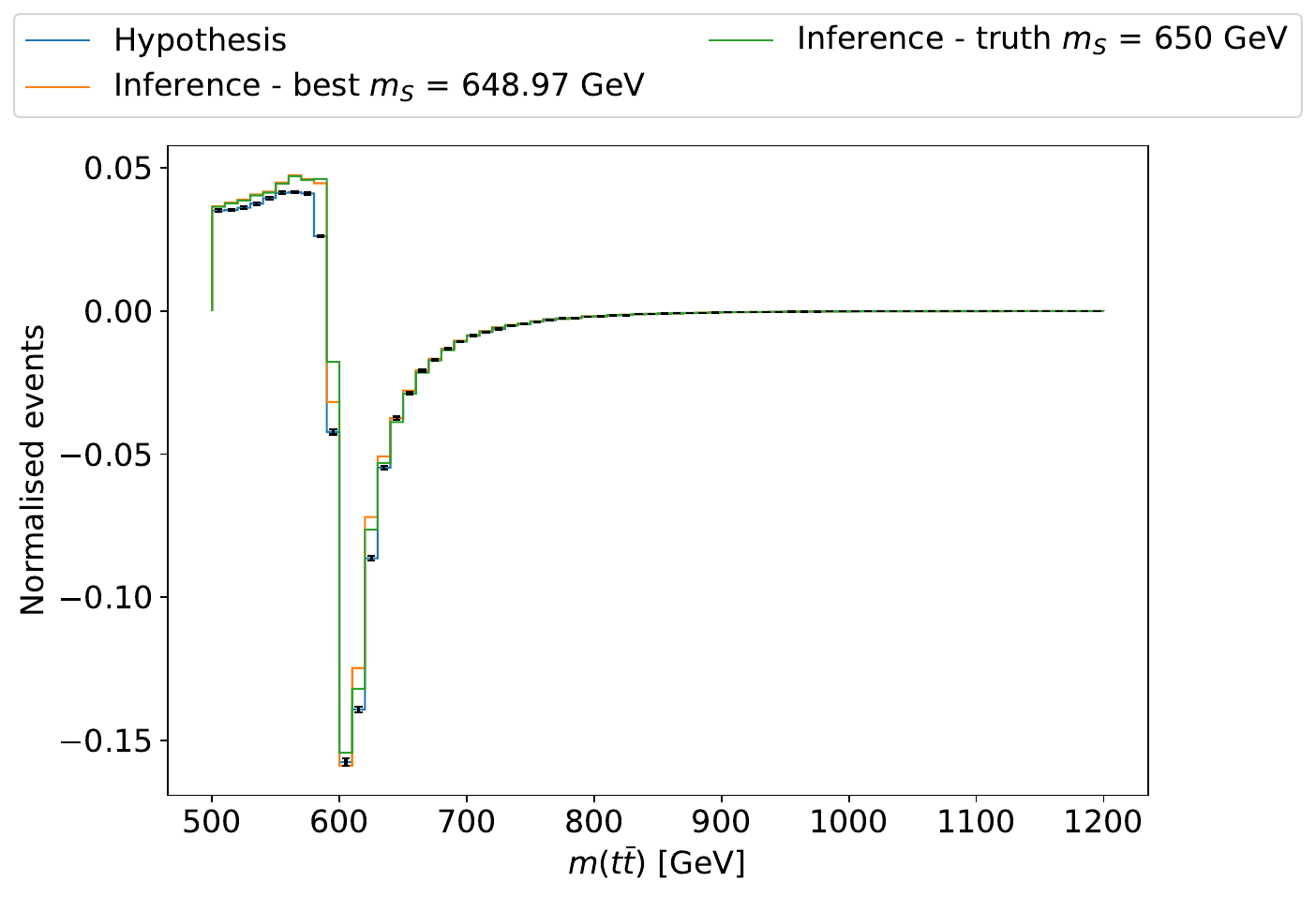}}
    \hfill
    \subfigure[]{\includegraphics[width=.48\linewidth]{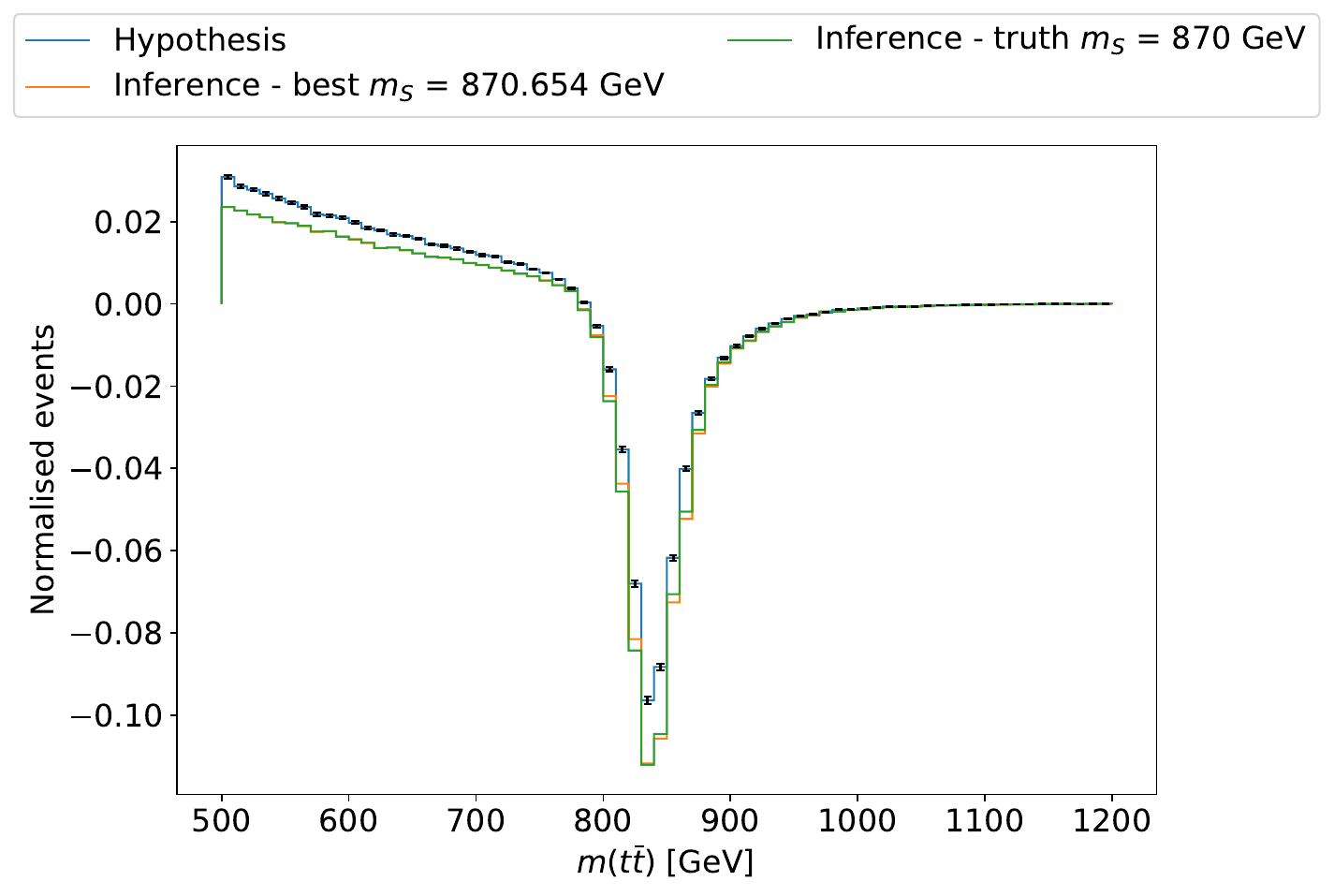}}
    \caption{Inference pipeline summary for (a) $650$ and (b) $870$ GeV resonance masses. In each figure, the blue line shows the truth hypothesis $\mttbar$ distribution, the orange line shows the best-fit parameter estimated via the inference pipeline, and the green line shows the distribution generated from performing inference with the truth value for the $m_S$ parameter. All distributions are normalised to unity.\label{fig:summary_1D_mass}}
\end{figure}

%%%%%%%%%%%%%%%%%%%%%%%%%%%%%%%%%%%%%%%%%%%%%%
To test the inference pipeline, we use hypothetical data from each of the holdout mass hypotheses shown in Tab.~\ref{tab:parameters} ($m_S = 650, 870~\text{GeV}$). The shapes of the $L^2$ scans for $m_S \in [550,1100]$ GeV can be seen in Fig.~\ref{fig:scan_1D_mass}, with the truth value marked by the green circle and the inference result with the red cross. We observe a clear minimum for the $L^2$ statistic. Figure~\ref{fig:summary_1D_mass} presents the comparison between the shapes of the hypothesis, the shape given by our RoSMM model at the estimated $m_S$, and the shape given by the RoSMM model at the truth $m_S$. The performance is comparable to the 1D study with fixed $m_S$; in particular, the parameter estimation reproduces the truth mass accurately.

%%%%%%%%%%%%%%%%%%%%%%%%%%%%%%%%%%%%%%%%%%%%%%
%%%%%%%%%%%%%%%%%%%%%%%%%%%%%%%%%%%%%%%%%%%%%%

%%%%%%%%%%%%%%%%%%%%%%%%%%%%%%%%%%%%%%%%%%%%%%
\begin{figure}[!t]
    \centering
    \subfigure[\label{fig:scan_2Da}]{\includegraphics[width=.48\linewidth]{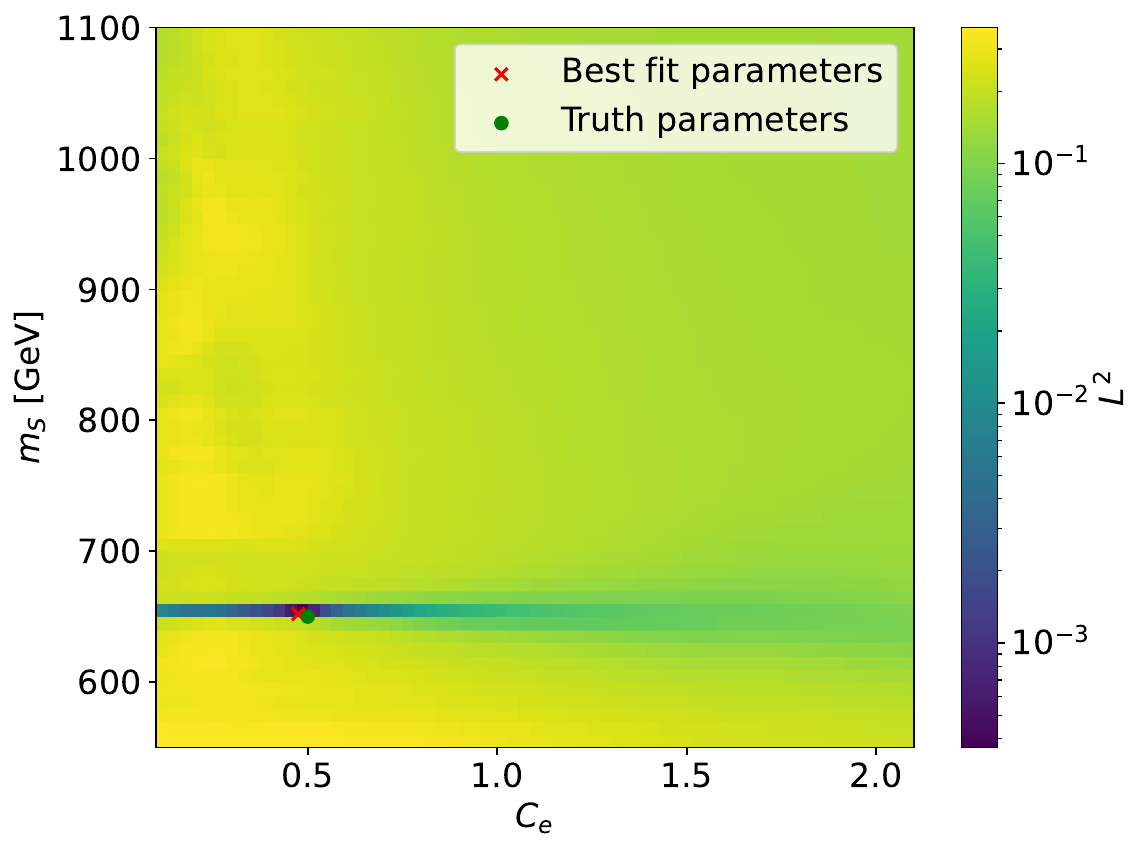}}
    \hfill
    \subfigure[\label{fig:scan_2Db}]{\includegraphics[width=.48\linewidth]{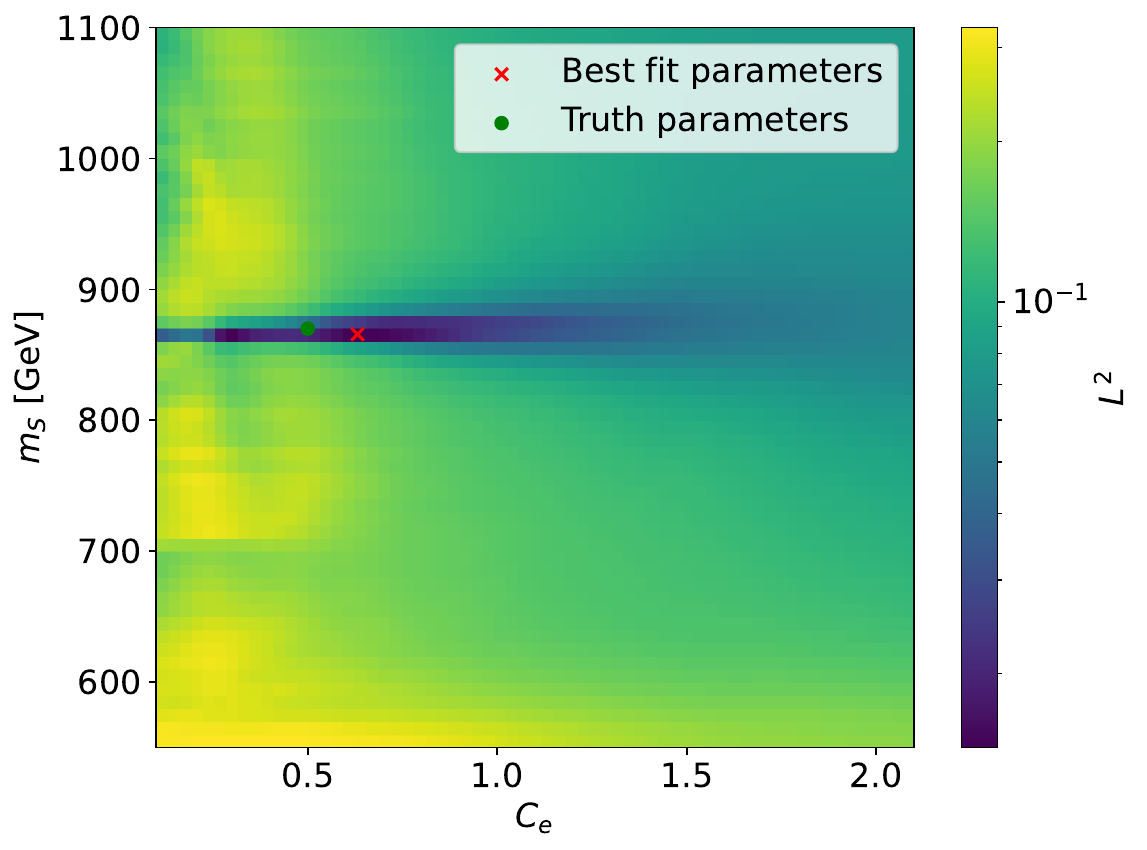}}
    \hfill
    \subfigure[\label{fig:scan_2Dc}]{\includegraphics[width=.48\linewidth]{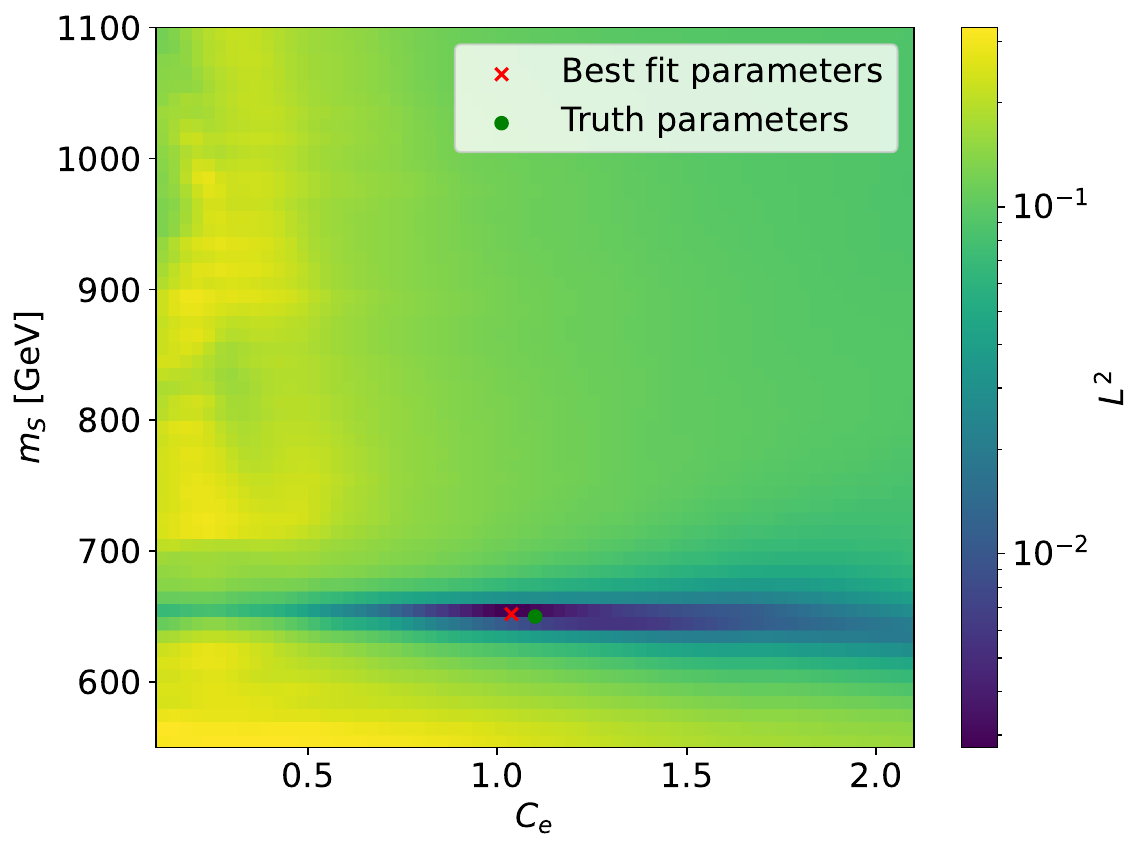}}
    \hfill
    \subfigure[\label{fig:scan_2Dd}]{\includegraphics[width=.48\linewidth]{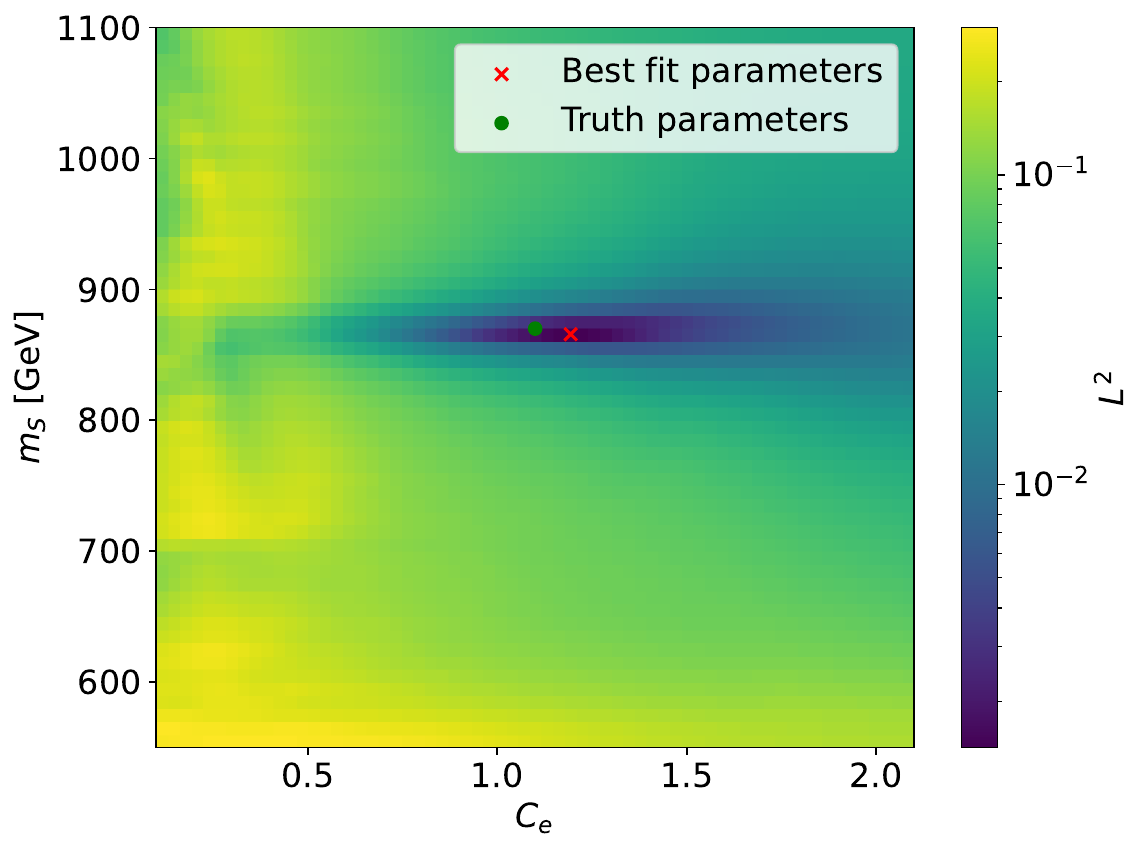}}
    \caption{$L^2$ scans heat map for $(C_e,m_S/\text{GeV})$ values (a) $(0.5,650)$, (b) $(0.5,870)$, (c) $(1.1,650)$, and (d) $(1.1,870)$ hypothesis. The green circle shows the truth $(C_e,m_S)$ value, and the red cross shows the estimated value via the inference pipeline.\label{fig:scan_2D}}
\end{figure}
%%%%%%%%%%%%%%%%%%%%%%%%%%%%%%%%%%%%%%%%%%%%%%

Overall, the $m_S$ parameter estimation has a bias of $\approx 0.2\%$. According to our statistical uncertainty estimation procedure, this cannot be explained solely by statistical MC uncertainty. The bias of the extraction of $C_e$ described in the previous Sec.~\ref{sec:ce} is $\approx 5\%$ at low $C_e$ values, and, again, cannot be explained by statistical fluctuations alone. For higher $C_e$ values, the estimate is unbiased, and the differences are within our statistical uncertainty. 
This alludes to interpolation biases induced by the coarse-graining of our training set. Such effects can be combated through a more finely grained hypotheses mesh, which will also be more relevant when additional final states' kinematics are included in fully-hadronised final states~\cite{forth}.

%%%%%%%%%%%%%%%%%%%%%%%%%%%%%%%%%%%%%%%%%%%%%%
\subsection{Inference of the full toy model}
\label{sec:res2d}
%%%%%%%%%%%%%%%%%%%%%%%%%%%%%%%%%%%%%%%%%%%%%%
With confidence that an accurate inference is possible for the model's parameters individually, we turn to estimating $C_e$ and $m_S$ simultaneously. The dataset is formed of 854k class-0 events. The total number of class-1 events is 54.1m. The mass values used for training and inference validation (holdout values) are shown in Tab.~\ref{tab:parameters}.

To test the inference pipeline, we use hypothetical data from the following holdout hypothesis $(C_e, m_S /\text{GeV}) = [ (0.5, 650), (0.5, 870), (1.1, 650), (1.1, 870) ]$. The $L^2$ heat maps scans in a grid with 50 evenly spaced points with each parameter axis between the ranges $m_S \in [550,1100]$ GeV and $C_e=[0.1, 2.1]$ are displayed in Fig.~\ref{fig:scan_2D}. The truth value is marked by the green circle, and the inference result is marked with the red cross. As in the 1D studies, we also observe a minimum in the $L^2$ statistic for the 2D case.  

\begin{figure}[!t]
    \centering
    \subfigure[]{\includegraphics[width=.48\linewidth]{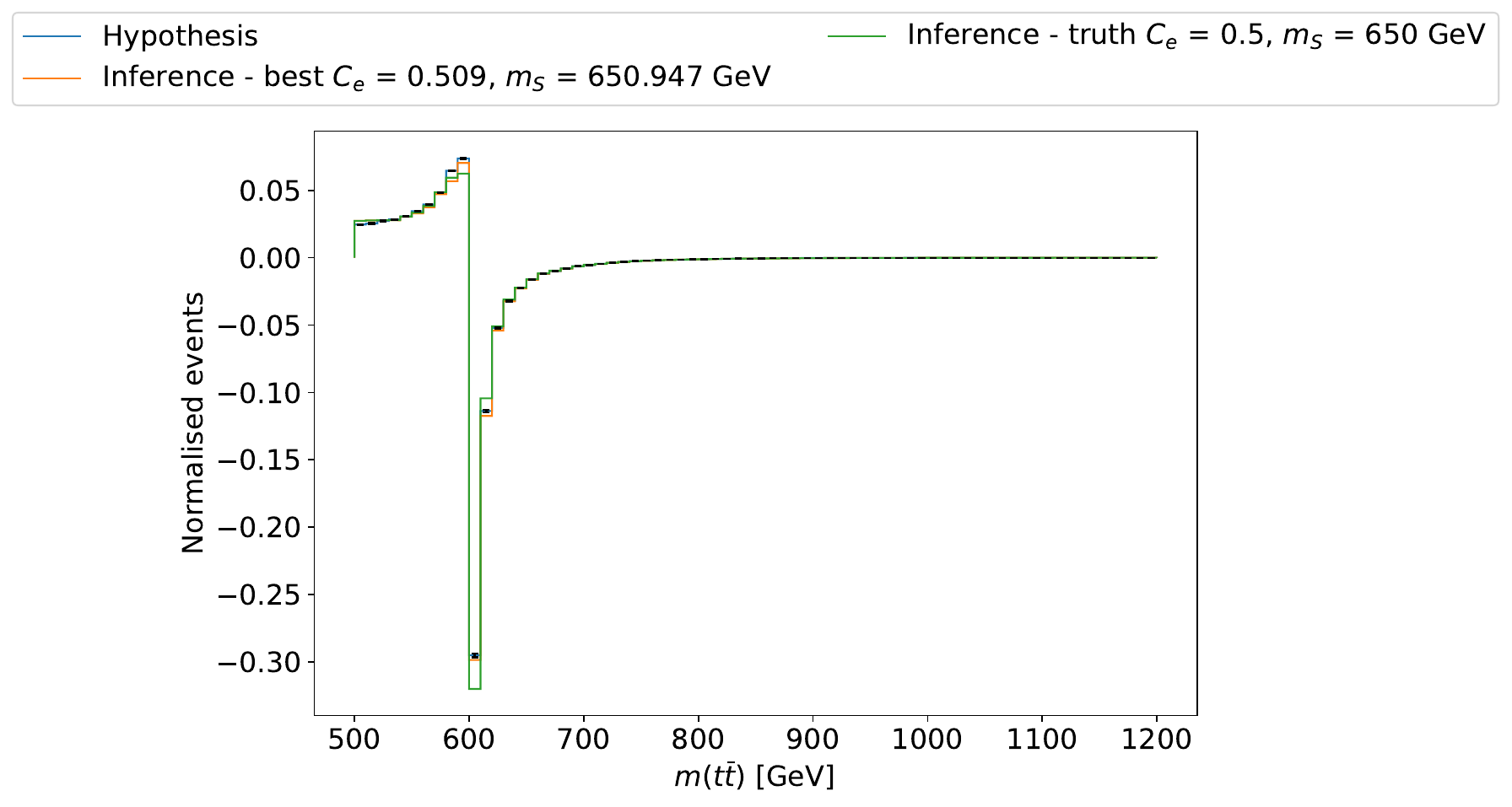}}
    \hfill
    \subfigure[]{\includegraphics[width=.48\linewidth]{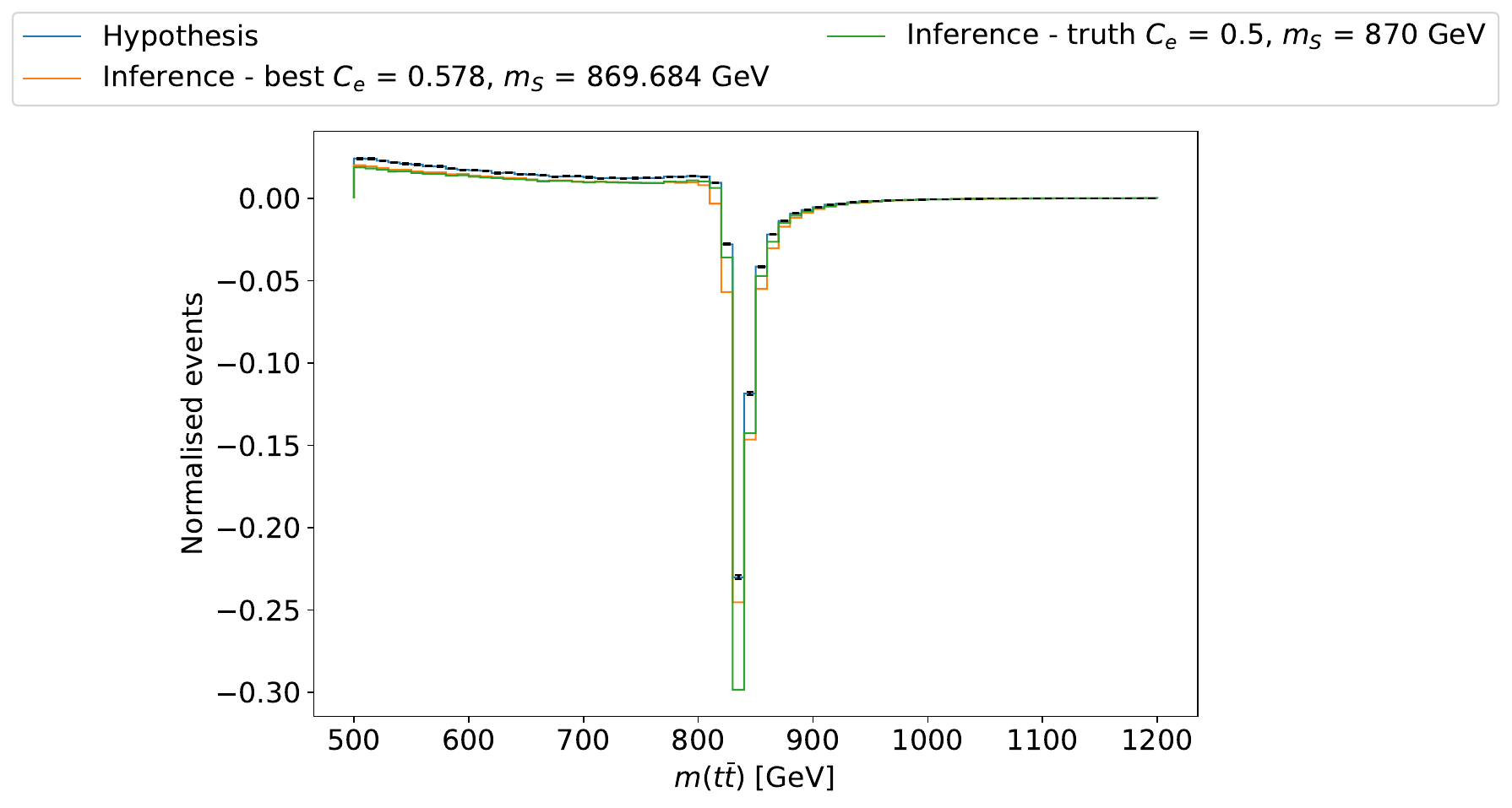}}
    \hfill
    \subfigure[]{\includegraphics[width=.48\linewidth]{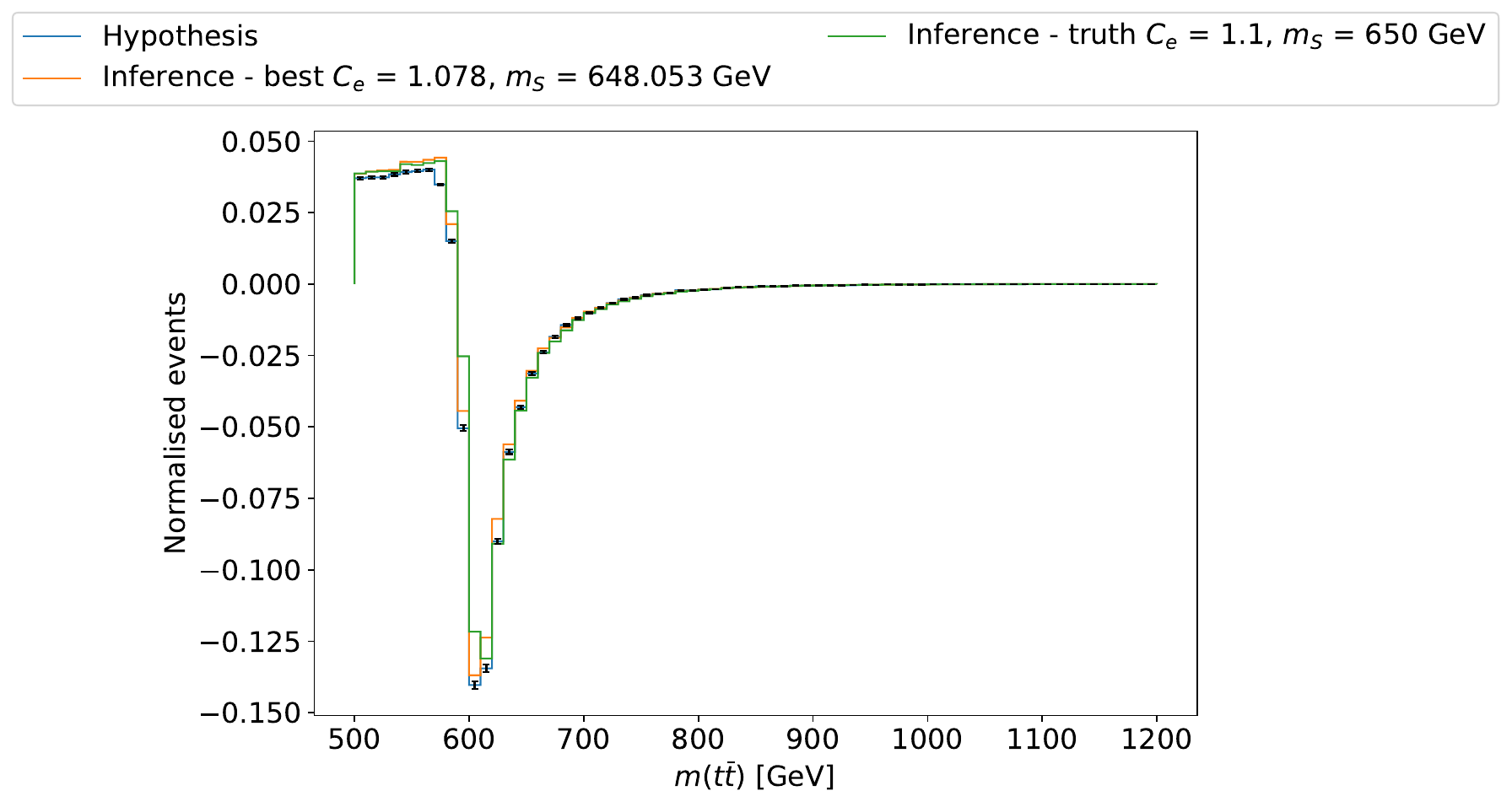}}
    \hfill
    \subfigure[]{\includegraphics[width=.48\linewidth]{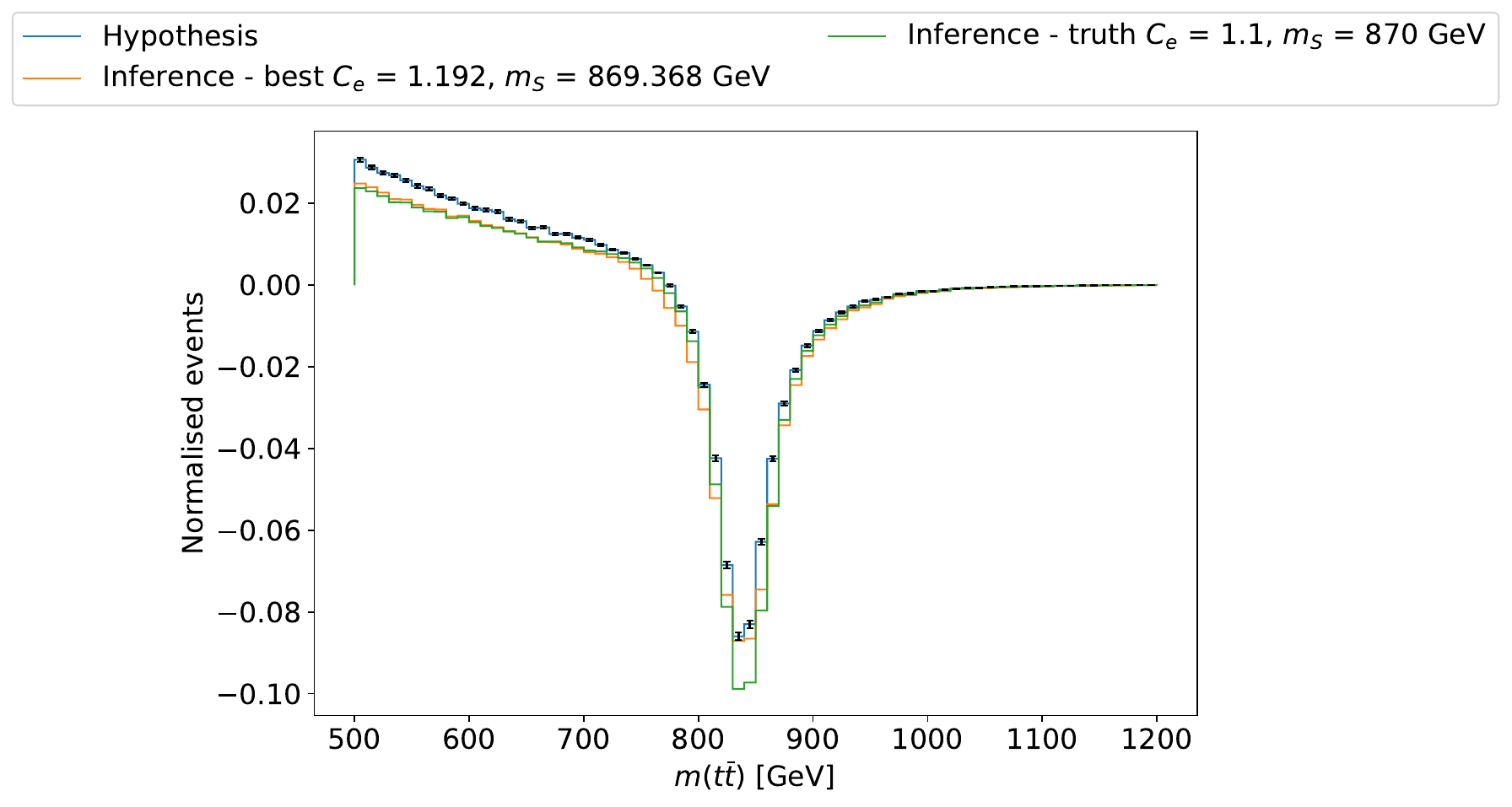}}
    \caption{Inference pipeline summary for values $(C_e,m_S/\text{GeV})$ (a) $(0.5,650)$, (b) $(0.5,870)$, (c) $(1.1,650)$, and (d) $(1.1,870)$. In each plot, the blue line shows the truth hypothesis $\mttbar$ distribution, the orange line shows the best-fit parameter estimated via the inference pipeline, and the green line shows the distribution generated from performing inference with the truth value for the $(C_e,m_S)$ parameters. All the distributions are normalised to unity.\label{fig:summary_2D}}
\end{figure}

Figure~\ref{fig:summary_2D} presents the comparison between the shapes of the hypothesis, the shape given by our RoSMM model at the estimated $m_S$, and the shape given by the RoSMM model at the truth $m_S$. We perform 10k pseudo-experiments, repeating the $L^2$ scans to obtain the statistical uncertainty band. These results are consistent with the extraction of $C_e,m_S$ individually, described in Secs.~\ref{sec:ce} and~\ref{sec:ms}.

%
%%%%%%%%%%%%%%%%%%%%%%%%%%%%%%%%%%%%%%%%%%%%%%
%%%%%%%%%%%%%%%%%%%%%%%%%%%%%%%%%%%%%%%%%%%%%%
%
%
In Figures~\ref{fig:scan_2D} we can see how the RoSMM inference pipeline produces an $L^2$ landscape that constrains the $m_S$ parameter much better than $C_e$. This is compatible with the shapes of the $L^2$ curves presented in Figures~\ref{fig:scan_1D_coupling}~and~\ref{fig:scan_1D_mass}. As anticipated from our 1D results, the model contains (phenomenologically irrelevant) biases beyond statistical uncertainties of the observed hypothetical samples.

%%%%%%%%%%%%%%%%%%%%%%%%%%%%%%%%%%%%%%%%%%%%%%
\begin{figure}
    \centering
    \subfigure[]{\includegraphics[width=.48\linewidth]{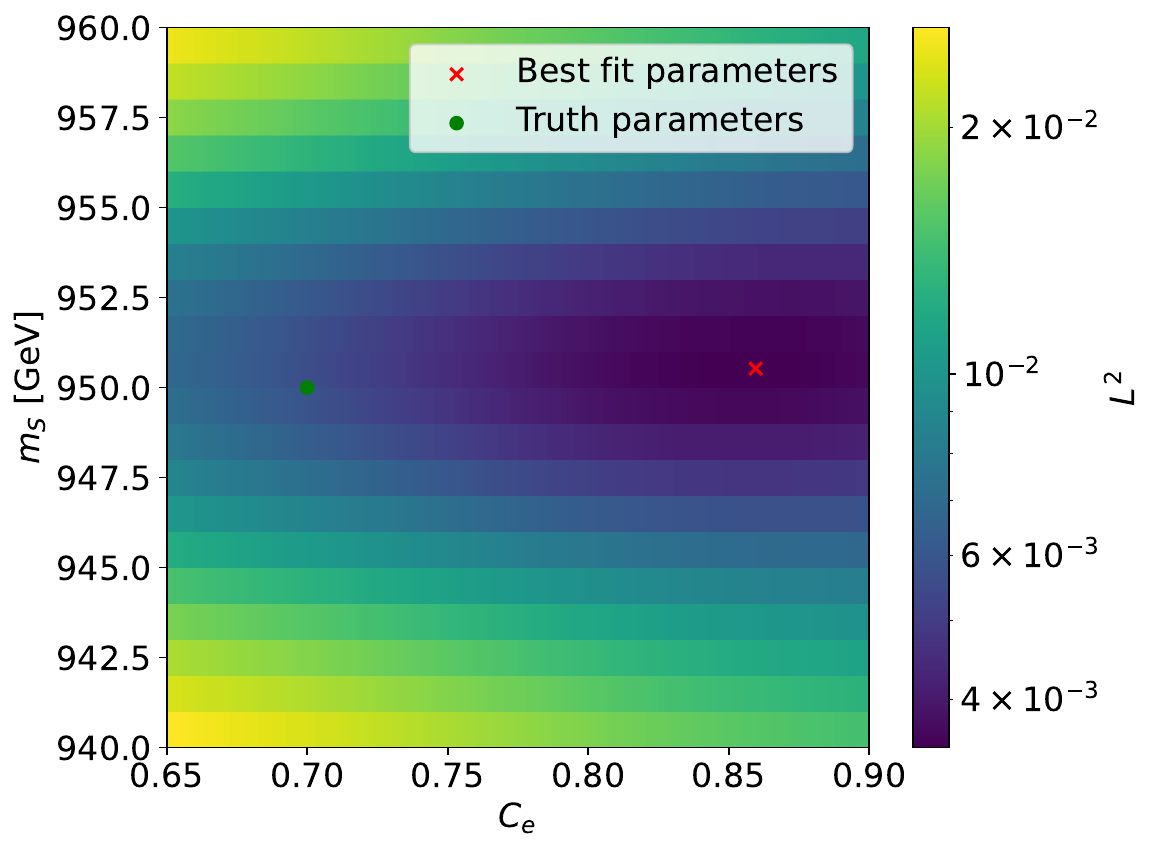}}
    \hfill
    \subfigure[]{\includegraphics[width=.48\linewidth]{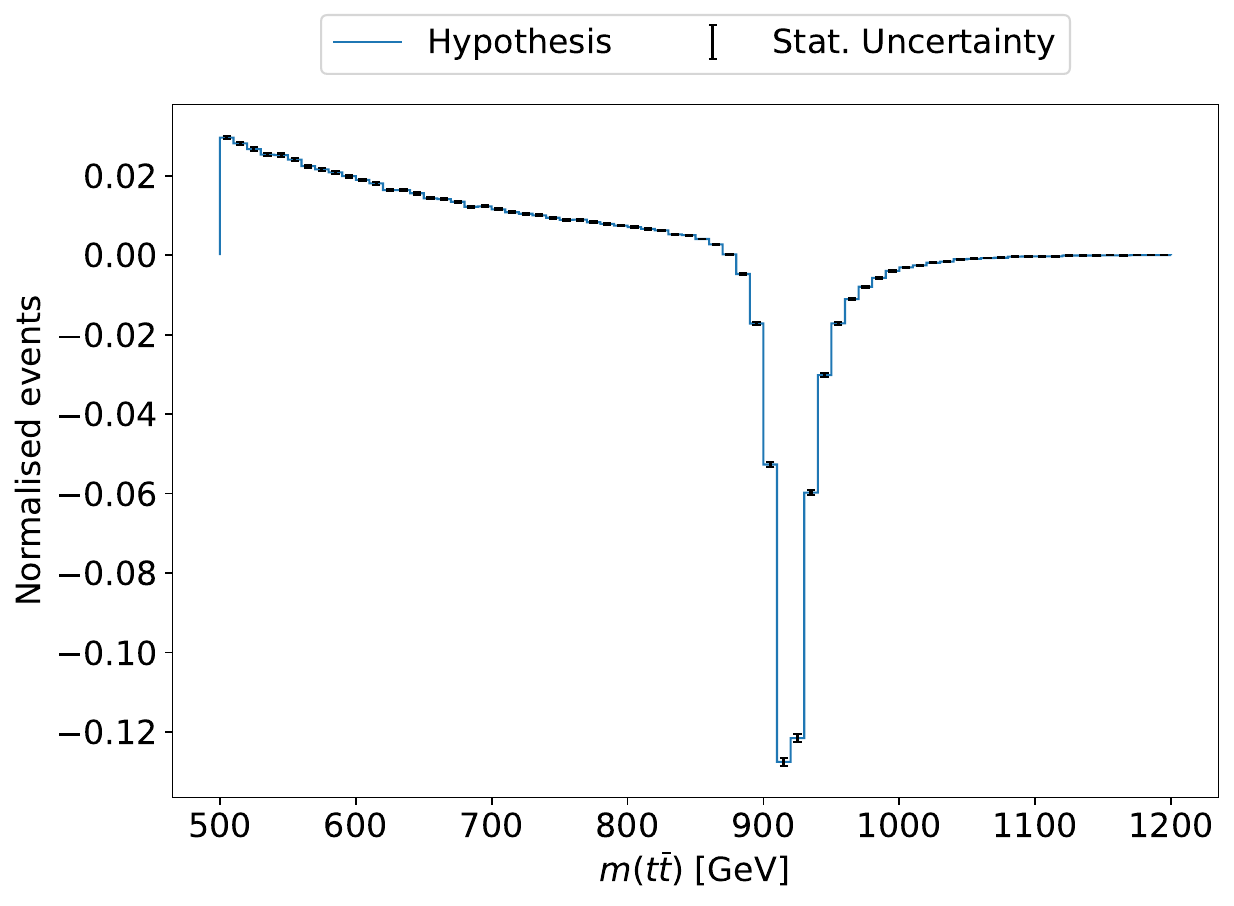}}
    \hfill
    \subfigure[]{\includegraphics[width=.58\linewidth]{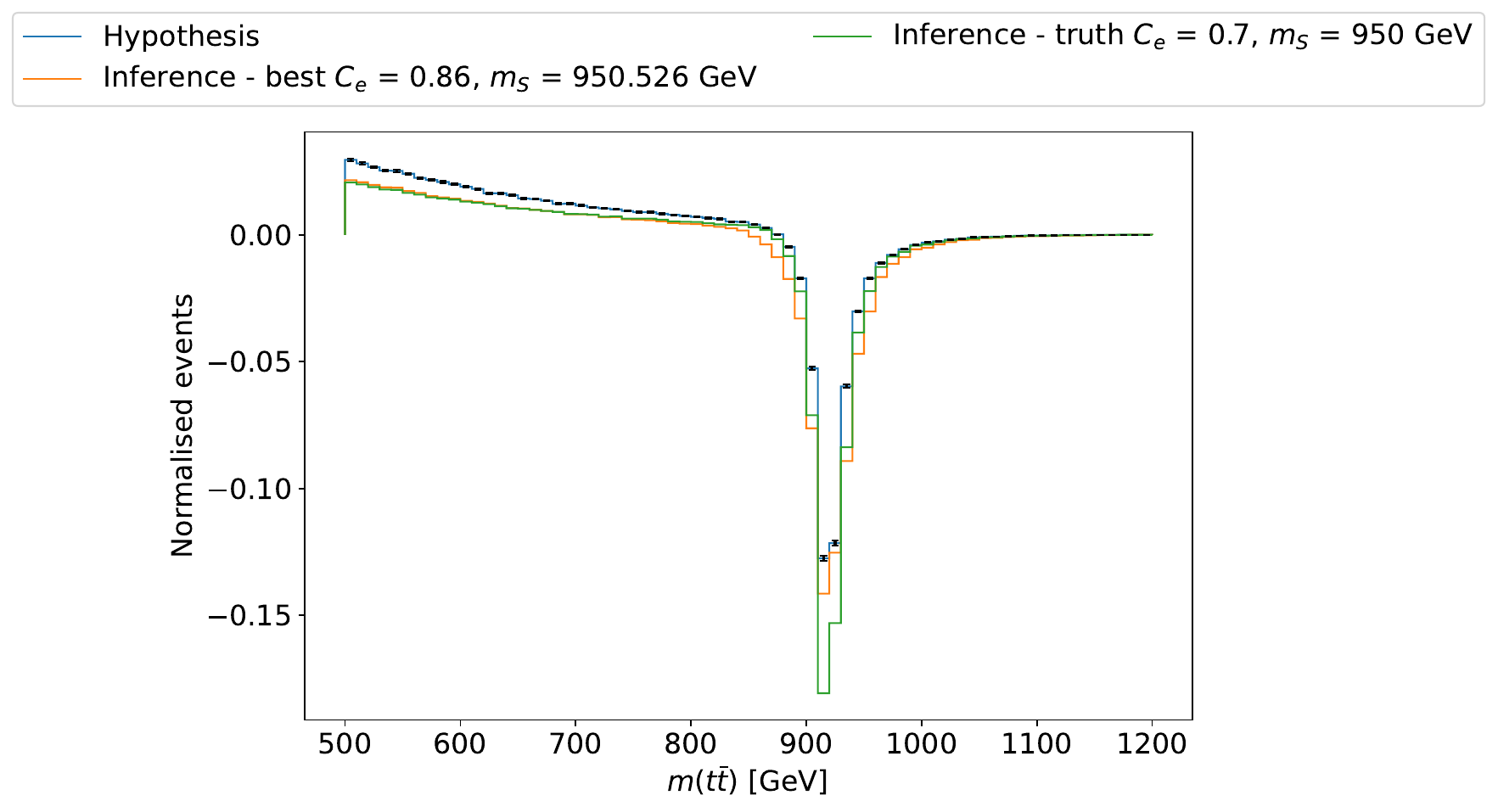}}
    \caption{Inference review for $(C_e=0.7,m_S=950)$. The $L^2$ heat map can be seen in (a). The hypothesis shape is shown in (b). The inference summary is seen in (c).\label{fig:width_test_from_model}}
\end{figure}
%%%%%%%%%%%%%%%%%%%%%%%%%%%%%%%%%%%%%%%%%%%%%%
%%%%%%%%%%%%%%%%%%%%%%%%%%%%%%%%%%%%%%%%%%%%%%
\begin{figure}
    \centering
    \subfigure[]{\includegraphics[width=.48\linewidth]{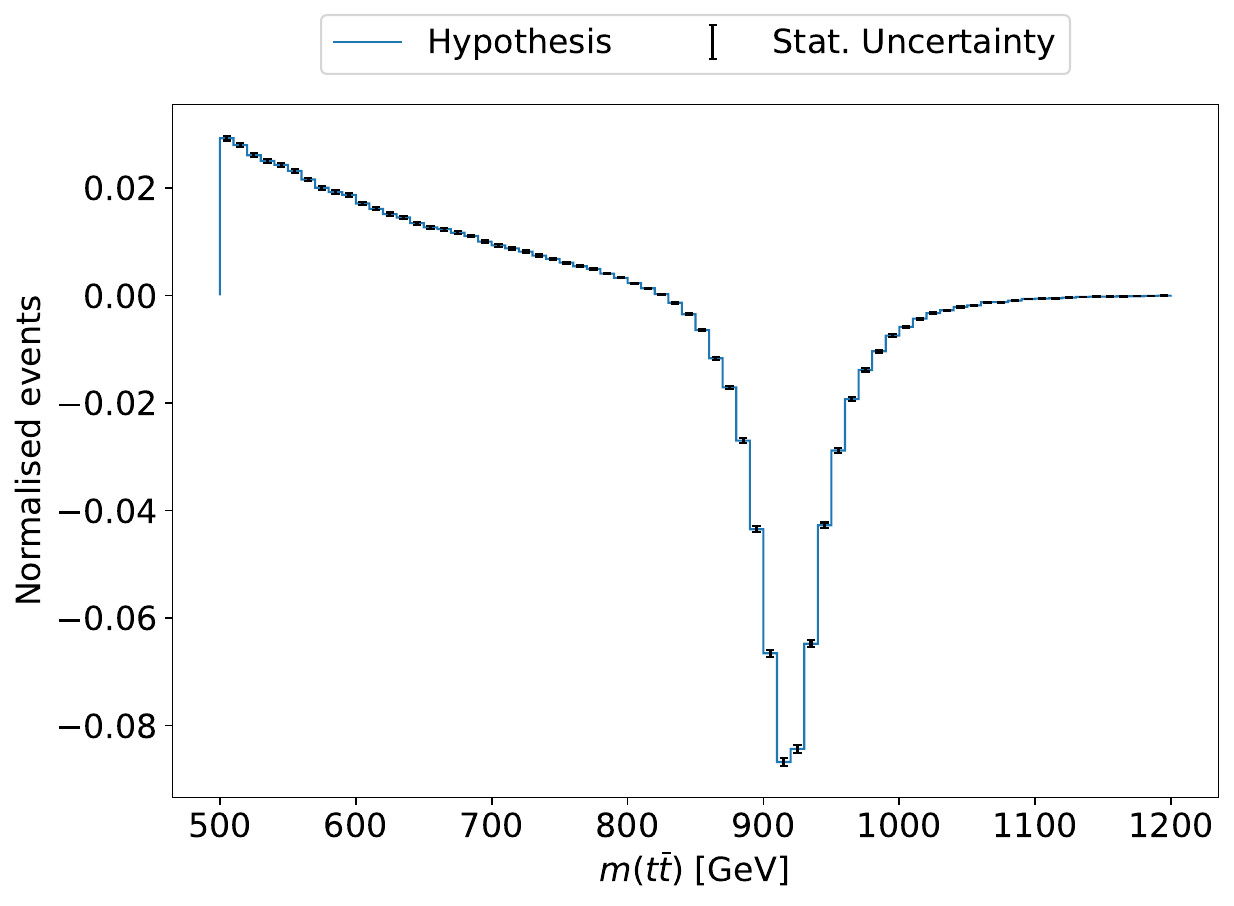}}
    \hfill
    \subfigure[]{\includegraphics[width=.48\linewidth]{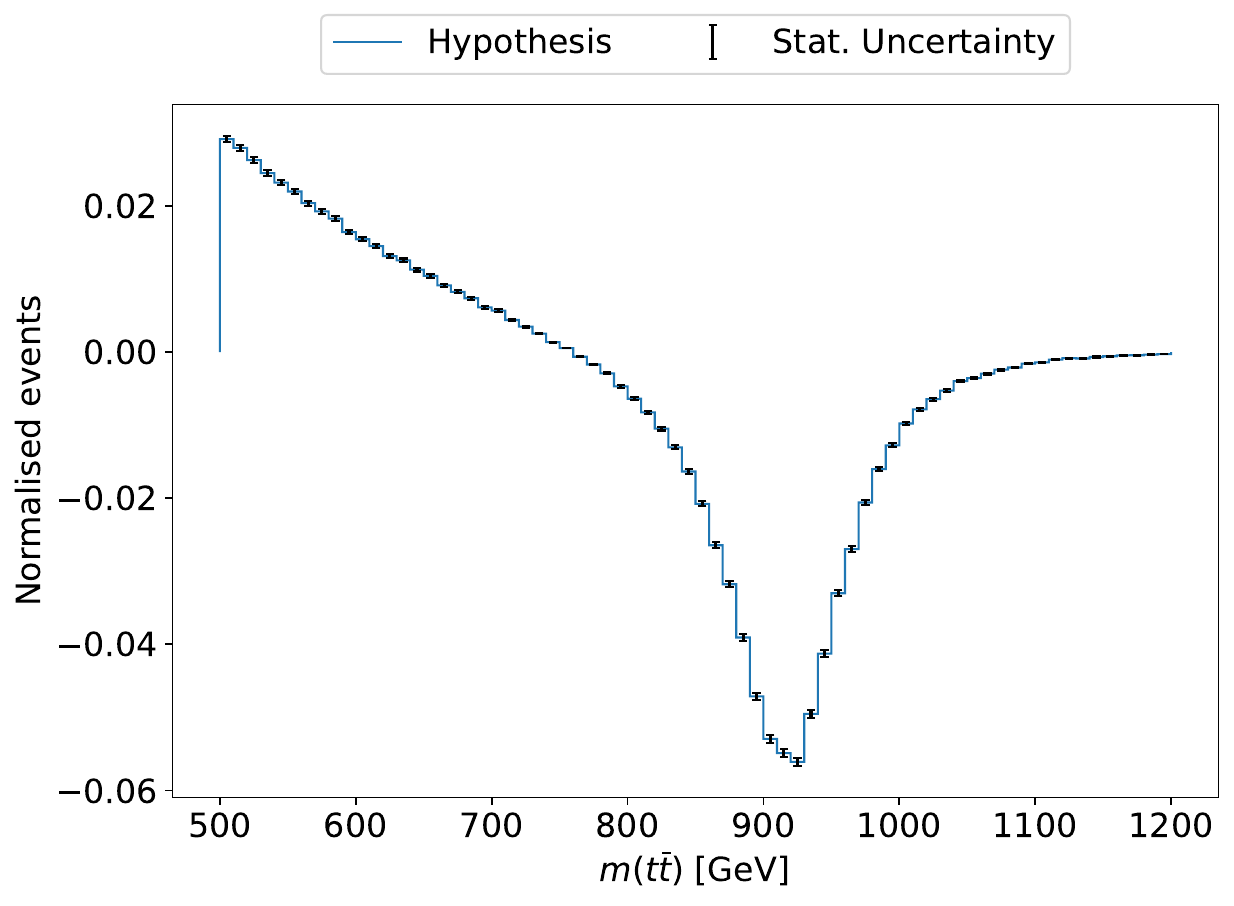}}
    \hfill
    \subfigure[]{\includegraphics[width=.48\linewidth]{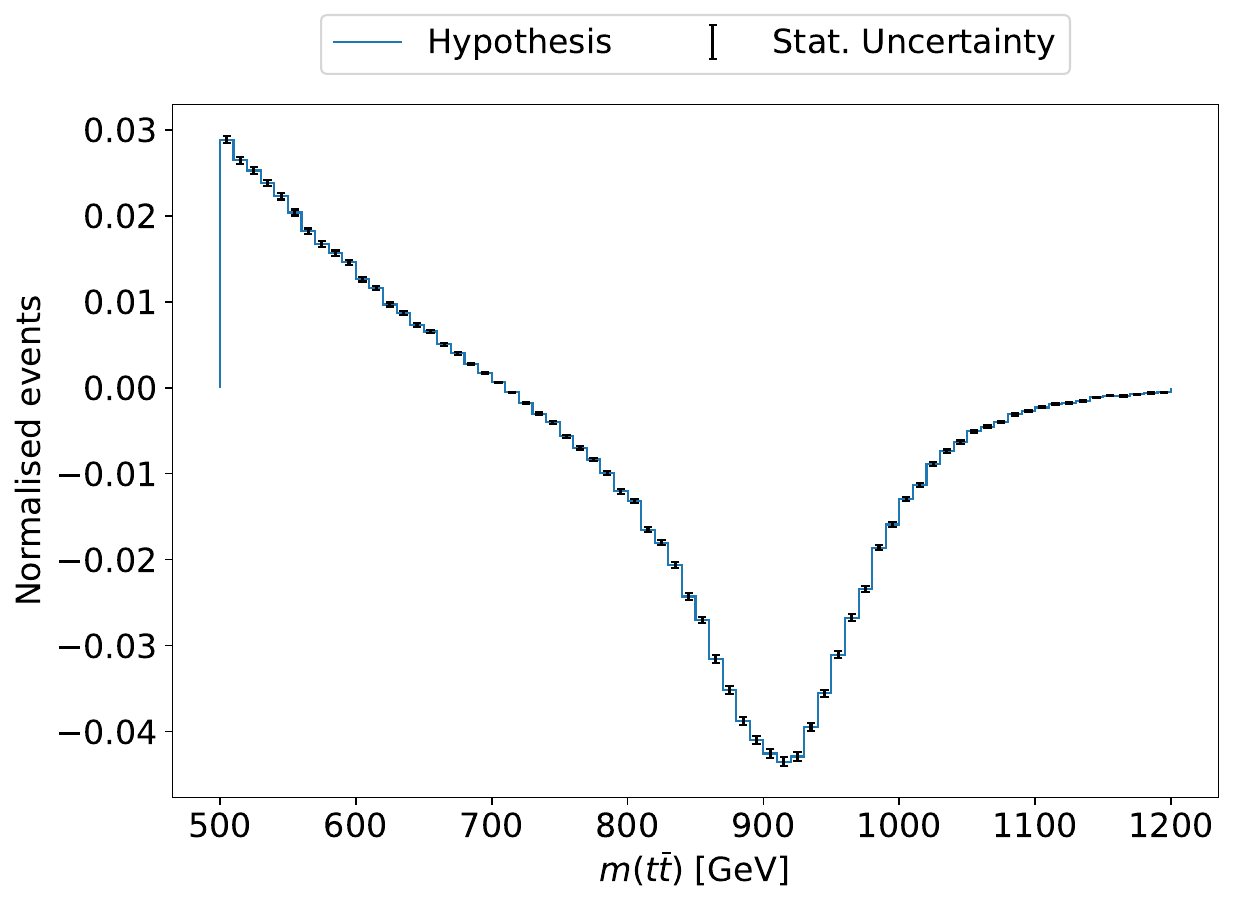}}
    \hfill
    \subfigure[]{\includegraphics[width=.48\linewidth]{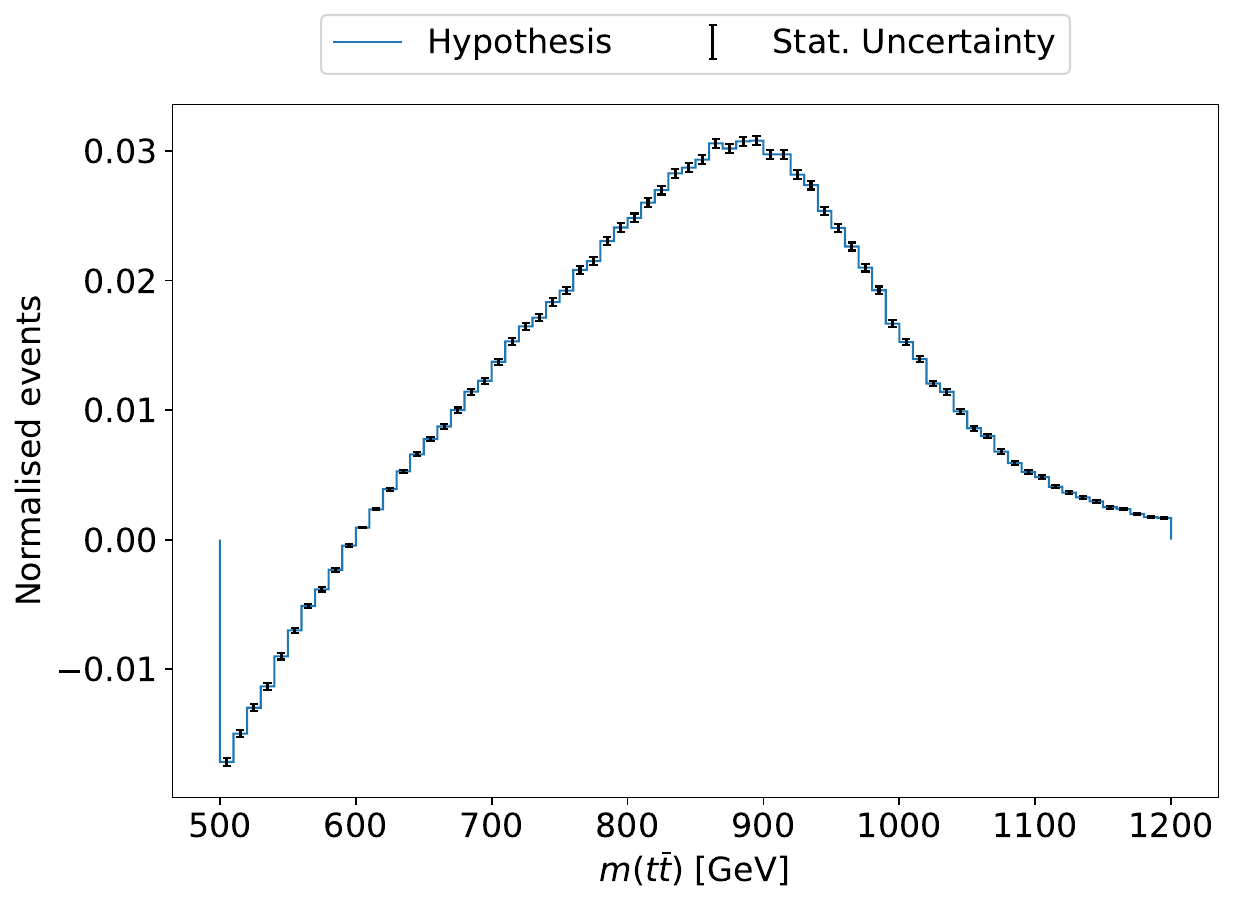}}
    \caption{Hypothesis $\mttbar$ distribution for different values of $\Gamma_S$ as a percentage of the $m_S=950$ GeV parameter. (a) shows 5\%, (b) 10\%, (c) 15\%, and (d) 30\%.\label{fig:width_test_manual}}
\end{figure}
%%%%%%%%%%%%%%%%%%%%%%%%%%%%%%%%%%%%%%%%%%%%%%

%%%%%%%%%%%%%%%%%%%%%%%%%%%%%%%%%%%%%%%%%%%%%%
\subsection{Going beyond: Relaxing model constraints}
\label{sec:modeldependence}
%%%%%%%%%%%%%%%%%%%%%%%%%%%%%%%%%%%%%%%%%%%%%%
So far, we have limited ourselves to the fairly rigid model correlations that determine the entire phenomenology through the choice of the two parameters. This raises the question of whether dip-hunting in the form described here is limited to our specific model choice. To this end, we investigate the effect of varying the resonance width to assess how the inference pipeline captures these changes. This effectively decouples production of the scalar from its decay. In more traditional bump-hunts, this approach directly informs (pseudo-)observables such as production cross section and branching ratios. When interference is present, such an interpretation is not valid. 

For a given $C_e$ and $m_S$ value, the width of the resonance in our model is derived via Eq.~\eqref{eq:4.1}. As our performance baseline, we generate a sample with $C_e = 0.7$ and $m_S=950$ GeV. The width derived from Eq.~\eqref{eq:4.1} for these parameter values is $\Gamma_S\approx22$ GeV, which corresponds to $\approx 2.3\%$ of $m_S$. When we apply the inference pipeline to this sample, we obtain the results seen in Fig.~\ref{fig:width_test_from_model}. These values are qualitatively consistent with the $C_e$ bias ($\approx$ 15\%) observed for the closest hypothesis values studied in Fig.~\ref{fig:scan_2Db}.

%%%%%%%%%%%%%%%%%%%%%%%%%%%%%%%%%%%%%%%%%%%%%%
\begin{figure}[!t]
    \centering
    \subfigure[\label{fig:width_test_summarya}]{\includegraphics[width=.48\linewidth]{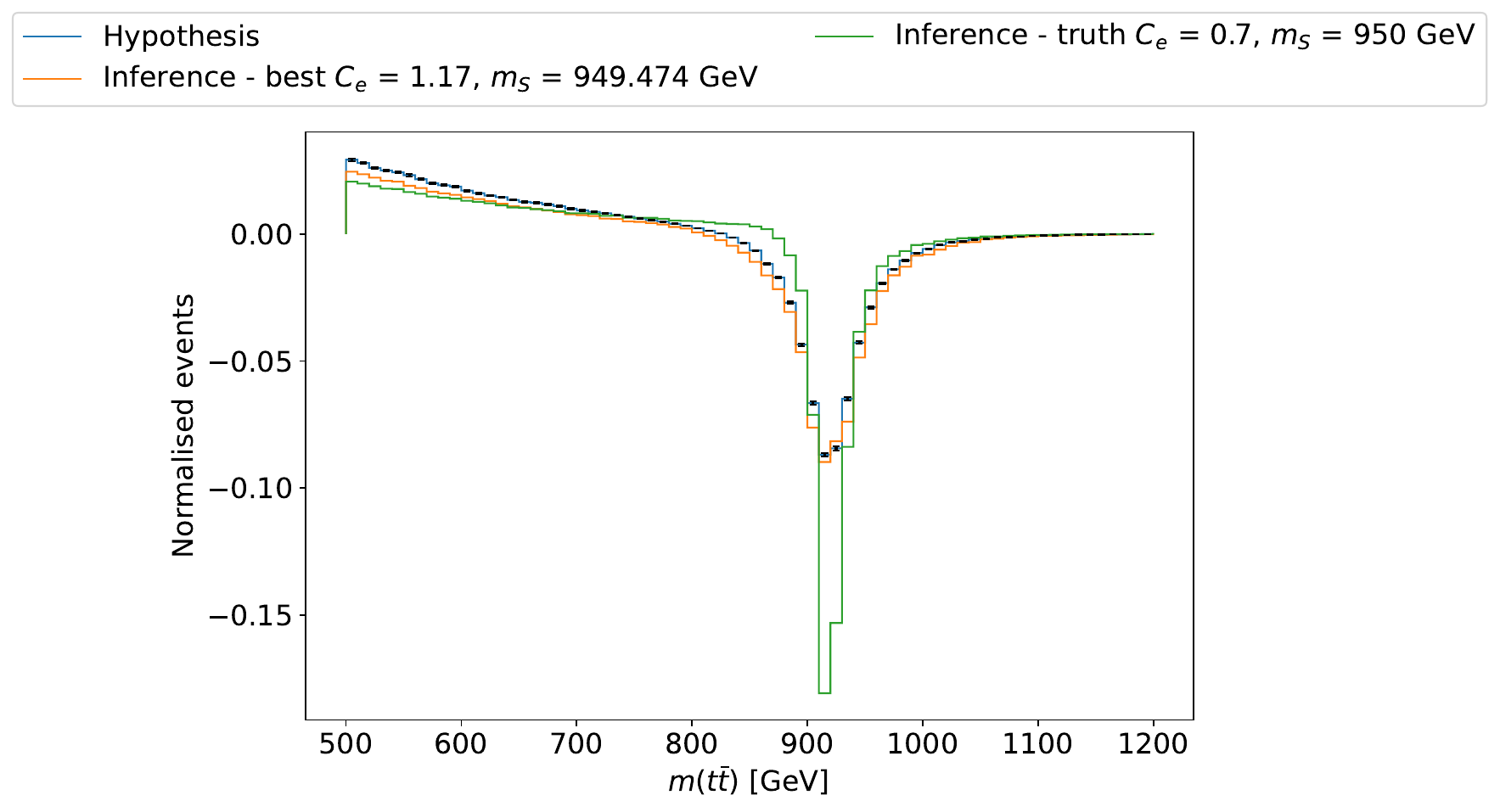}}
    \hfill
    \subfigure[\label{fig:width_test_summaryb}]{\includegraphics[width=.48\linewidth]{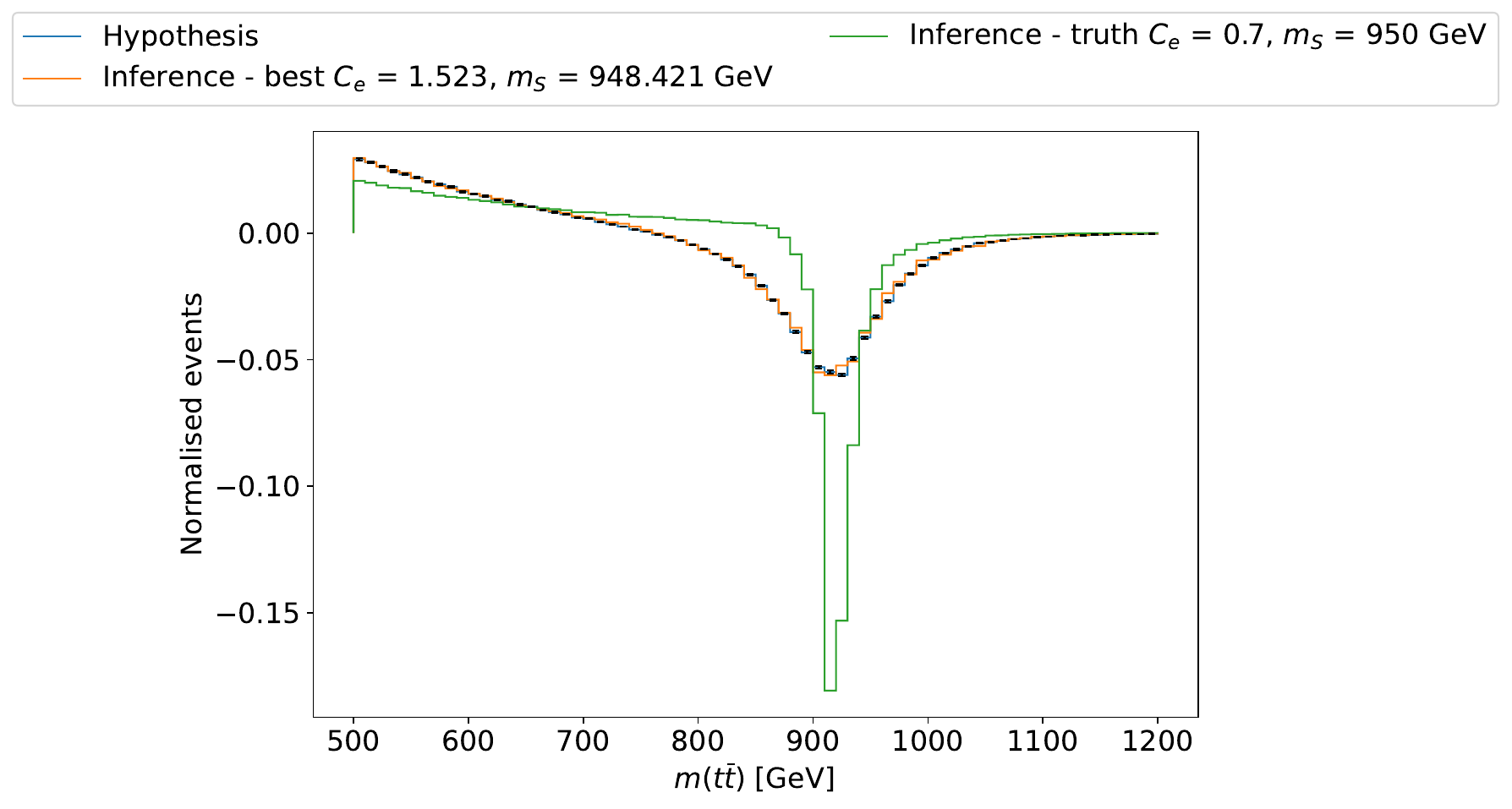}}
    \hfill
    \subfigure[\label{fig:width_test_summaryc}]{\includegraphics[width=.48\linewidth]{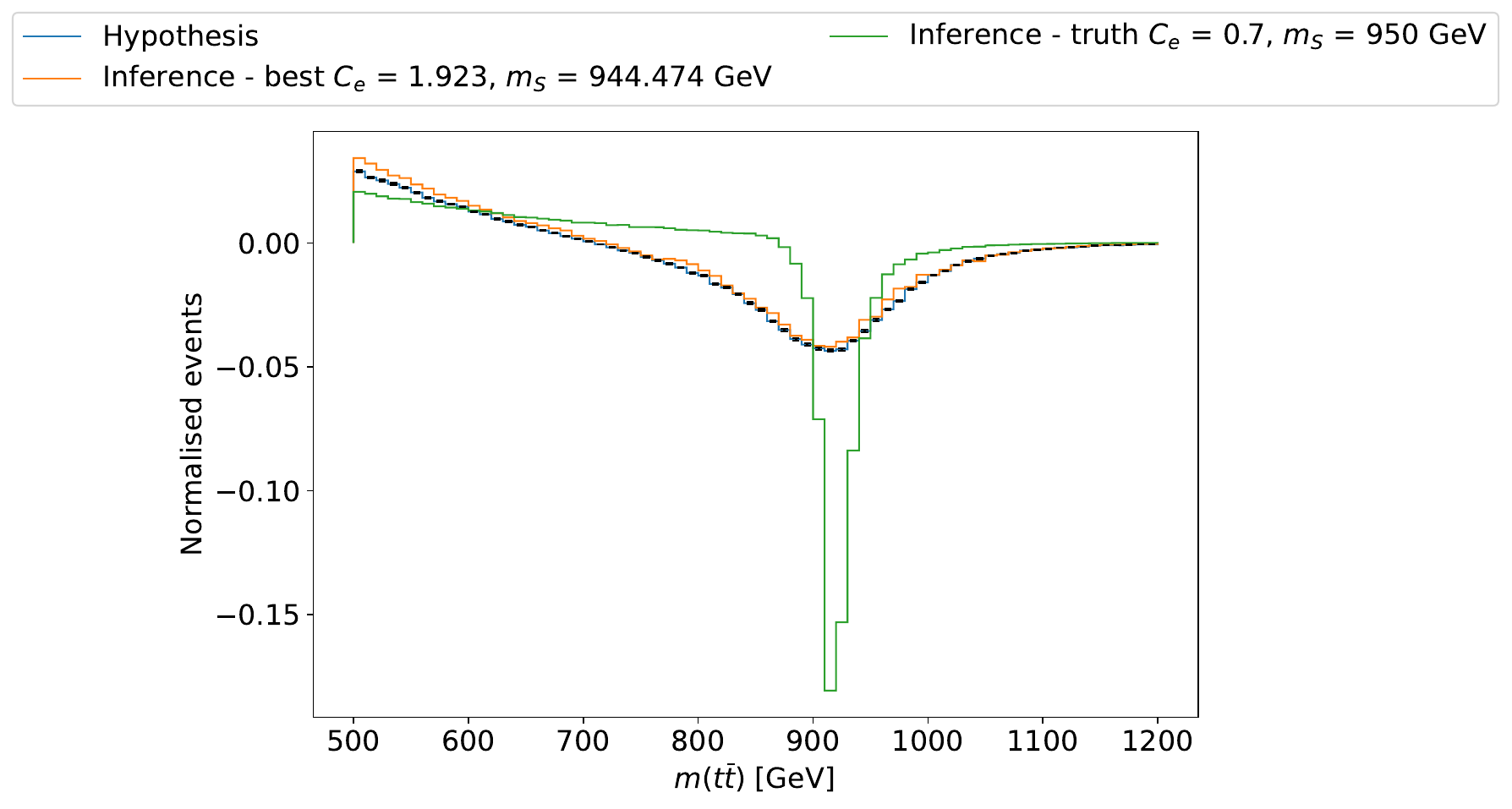}}
    \hfill
    \subfigure[\label{fig:width_test_summaryd}]{\includegraphics[width=.48\linewidth]{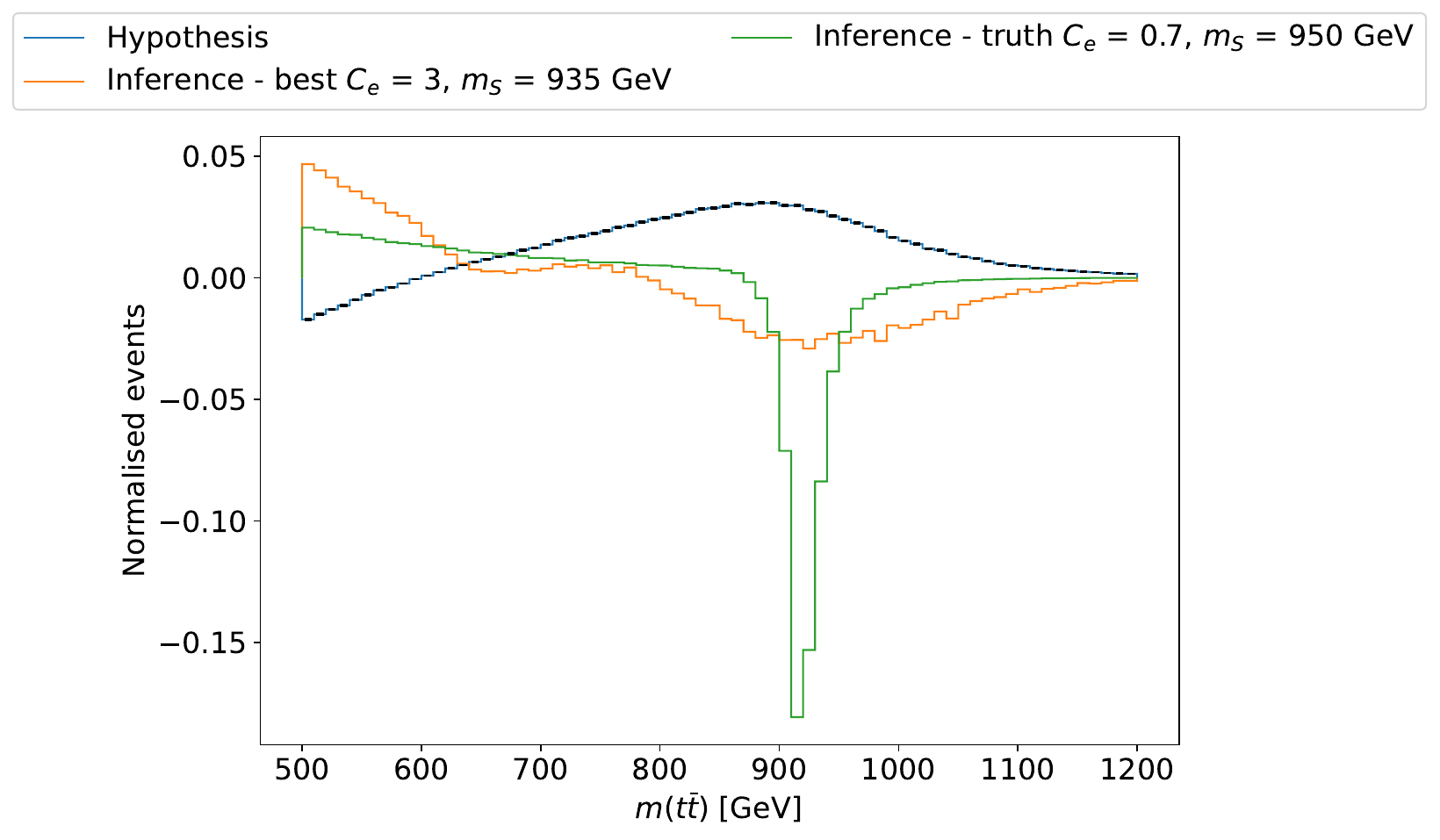}}
    \caption{Inference summary for different values of $\Gamma_S$ as a percentage of the $m_S=950$ GeV parameter. (a) shows 5\%, (b) 10\%, (c) 15\%, and (d) 30\%. The blue line shows the width-varying hypothesis, the orange line is the shape obtained from the inference pipeline, and the green line shows, for comparison, a reference where the width is not changed from the value obtained with Eq.~\eqref{eq:4.1}.\label{fig:width_test_summary}}
\end{figure}
%%%%%%%%%%%%%%%%%%%%%%%%%%%%%%%%%%%%%%%%%%%%%%

\begin{figure}
    \centering
    \subfigure[]{\includegraphics[width=.60\linewidth]{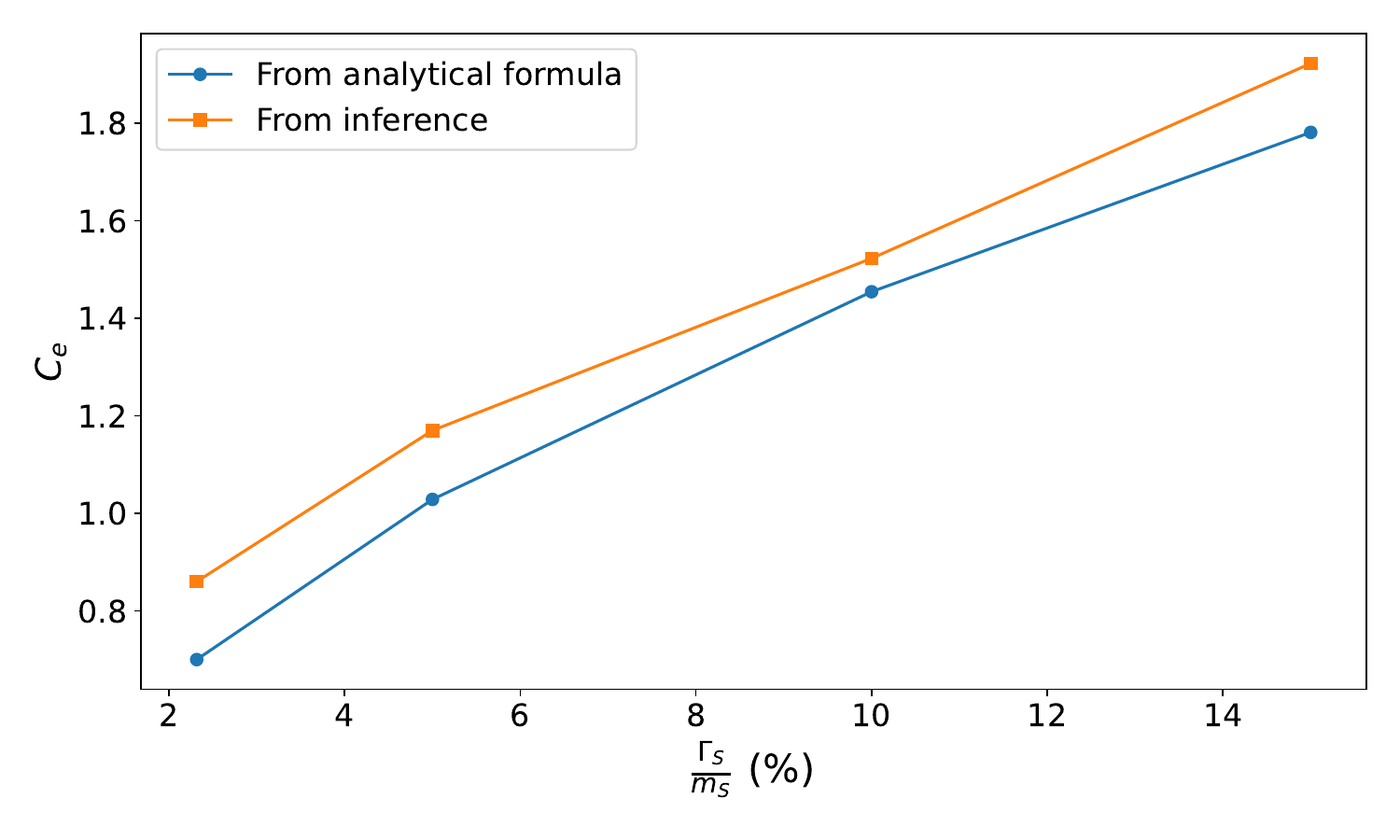}}
    \caption{$\Gamma_S$ dependence on $C_e$ according to Eq.~\eqref{eq:4.1}. The RoSMM model inference remaps the width-modified samples coupling $C_e=0.7$ to couplings that are compatible with the modified width up to 15\% ${\Gamma_S}/{m_S}$. As can be seen in Fig.~\ref{fig:width_test_summaryd}, our method shows significant deviations at ${\Gamma_S}/{m_S} = 30$\%. Hence, the ${\Gamma_S}/{m_S} = 30$\% point is not added to this figure.
    \label{fig:width_vs_coupling}}
\end{figure}

We also generate four additional hypotheses, but this time manually changing $\Gamma_S$ to values $\Gamma_S/m_S = (5, 10, 15, 30)\% $. $\Gamma_S/m_S=15\%$ for a mass of $650~\text{GeV}$ is starting to push the perturbative envelope with a coupling choice of $C_e\simeq 2$, with larger values quickly leading to an inconsistent phenomenological modelling (e.g. $C_e\simeq \pi$ for 30\%). The shapes of these hypotheses can be seen in Fig.~\ref{fig:width_test_manual}. The results of the inference scans for these widths can be seen in Fig.~\ref{fig:width_test_summary}. We observe that up to a width of 15\% $m_S$, the inference pipeline is able to find a shape compatible with the hypothesis, whilst not having been exposed to such large departures in the training. This is a consequence of the intrinsic degeneracy between $C_e$ and $\Gamma_S$. The quadratic relation shown in Eq.~\eqref{eq:4.1} between $C_e$ and $\Gamma_S$ can be seen in Fig.~\ref{fig:width_vs_coupling} as the blue line. The inference values from our pipeline can be seen as the orange line. We can see clearly how the inference ceases to work for larger $\Gamma_S$ values. This is expected because, as $\Gamma_S$ increases, the model needs to be able to perform inference with $C_e$ values well off the training grid. This can be seen in Fig.~\ref{fig:width_test_summaryc} at the depth of the distribution, where the inference results start to create noisy jumps in the $\mttbar$ bins. However, one can always extend the training to cover additional parts of the parameter space. Nonetheless, the ability of the network to still perform for large width modifications away from the top-correlation expressed in Eq.~\eqref{eq:4.1} demonstrates its applicability to a wide range of CP-even hypothesis theories. We stress that due to a different threshold behaviour, the CP-even coupling choice cannot represent effectively, e.g., a purely CP-odd state. Such a modification can in principle be incorporated, relatively independently of the method presented here, via an additional parameter that fits the relative contribution of $|{\cal{M}}_{S}|^2$ vs. $2\,\text{Re} \left( {\cal{M}}_{S}^\ast\, {\cal{M}}_{\text{con.},t} \right)$. This would then enable the extraction of the relatively larger CP-odd signal component expected compared to the CP-even interference correlation. Of course, mixed coupling choices lead to interpolating phenomenological consequences (see, e.g.,~\cite{Jung:2015gta}). We leave a more dedicated analysis of general scenarios for future work.

%%%%%%%%%%%%%%%%%%%%%%%%%%%%%%%%%%%%%%%%%%%%%%
%%%%%%%%%%%%%%%%%%%%%%%%%%%%%%%%%%%%%%%%%%%%%%

%%%%%%%%%%%%%%%%%%%%%%%%%%%%%%%%%%%%%%%%%%%%%%
\section{Conclusions}
\label{sec:conc}
%%%%%%%%%%%%%%%%%%%%%%%%%%%%%%%%%%%%%%%%%%%%%%
With searches for new interactions beyond the Standard Model well underway at the LHC and other experiments, BSM physics remains elusive. One possible avenue for relatively light new physics to evade direct detection is through accidental interference between the BSM signal and the SM background. These effects are present in a range of scalar extensions of the SM with top-philic properties, often with direct relevance for established BSM phenomena such as the observed matter-antimatter asymmetry~\cite{Basler:2019nas}. ATLAS and CMS include these effects in concrete analyses, e.g., of two-Higgs-doublet extensions~\cite{ATLAS:2025kmo,CMS:2025dzq}.

Contrary to well-established `bump-hunting' strategies, where the background is extrapolated into a signal region, and a signal `bump' extracted by background subtraction, interference-perturbed phenomenological outcomes do not lend themselves directly to such a procedure. The level of resonance distortion depends qualitatively and quantitatively on the interplay between the BSM signal and the interference effect, which in turn depends on the interplay between the signal and the background, yet in a quasi-analytically traceable way. This motivates applying machine learning techniques to predict and statistically validate the interference pattern observed in the data, given an expected background distribution. This approach, termed `dip-hunting', generalises `bump-hunting' to the interference-dominated final states and parameter regimes. We have developed this approach in a proof-of-principle investigation employing the Ratio of Signed Mixtures Model (RoSMM)~\cite{Drnevich:2024vfj} to learn a mapping between the SM  and the new physics contributions, demonstrating the dip-hunting capability in an application to top final states, which are particularly prone to such interference effects. Specifically, we use a real scalar simplified model extension that produces tightly correlated interference predictions given the model's parameters and find a robust extraction of the model parameters. This creates opportunities in two directions: Firstly, our approach opens the possibility of performing experimental analyses~\cite{ATLAS:2025kmo} with dramatically reduced overhead, thereby paving the way to model interference effects beyond the few benchmark scenarios typically considered and presented by the experiments. Secondly, given the robustness we observe in particular against modifying data correlations, within reasonable perturbative limits, we expect dip-hunting to still serve as a performant diagnostic of a potential interference-corrected particle threshold.\footnote{As always, the `correct' parameters of the underlying model, however, depend on the BSM model assumptions (just as is the case for bump-hunting).} This is explicitly demonstrated for CP-even resonances that do not follow the correlation pattern of our simple scalar extension. Our results, therefore, motivate further investigation of dip-hunting using more realistic final-state simulations and scalability (e.g., extending the methodology to signal-signal interference~\cite{Basler:2019nas,Bahl:2025you,Robens:2026spl}).

%%%%%%%%%%%%%%%%%%%%%%%%%%%%%%%%%%%%%%%%%%%%%%
\subsection*{Acknowledgements}
%%%%%%%%%%%%%%%%%%%%%%%%%%%%%%%%%%%%%%%%%%%%%%
Y.P. and D.A.B.M are supported by UK Research and Innovation [grant number EP/Z533865/1], the project was selected by the ERC, funded by UKRI. C.E. is indebted to Vulfpeck for inspiring the title of this paper.
%

%%%%%%%%%%%%%%%%%%%%%%%%%%%%%%%%%%%%%%%%%%%%%%
\section*{Data Availability Statement}
All the source code used for the results presented in this work is available via the Zenodo references~\cite{diego_baron_2026_19629894, diego_baron_2026_19629945}, linking to the relevant GitHub repositories.
%%%%%%%%%%%%%%%%%%%%%%%%%%%%%%%%%%%%%%%%%%%%%%

%%%%%%%%%%%%%%%%%%%
%\bibliographystyle{JHEP}
\bibliography{paper.bbl}

\providecommand{\href}[2]{#2}\begingroup\raggedright\begin{thebibliography}{10}

\bibitem{Gaemers:1984sj}
K.~J.~F. Gaemers and F.~Hoogeveen, {\it {Higgs Production and Decay Into Heavy
  Flavors With the Gluon Fusion Mechanism}},  {\em Phys. Lett. B} {\bf 146}
  (1984) 347--349.

\bibitem{Dicus:1994bm}
D.~Dicus, A.~Stange, and S.~Willenbrock, {\it {Higgs decay to top quarks at
  hadron colliders}},  {\em Phys. Lett. B} {\bf 333} (1994) 126--131,
  [\href{http://arxiv.org/abs/hep-ph/9404359}{{\tt hep-ph/9404359}}].

\bibitem{Basler:2019nas}
P.~Basler, S.~Dawson, C.~Englert, and M.~M{\"u}hlleitner, {\it {Di-Higgs boson
  peaks and top valleys: Interference effects in Higgs sector extensions}},
  {\em Phys. Rev. D} {\bf 101} (2020), no.~1 015019,
  [\href{http://arxiv.org/abs/1909.09987}{{\tt arXiv:1909.09987}}].

\bibitem{Jung:2015gta}
S.~Jung, J.~Song, and Y.~W. Yoon, {\it {Dip or nothingness of a Higgs resonance
  from the interference with a complex phase}},  {\em Phys. Rev. D} {\bf 92}
  (2015), no.~5 055009, [\href{http://arxiv.org/abs/1505.00291}{{\tt
  arXiv:1505.00291}}].

\bibitem{Frederix:2007gi}
R.~Frederix and F.~Maltoni, {\it {Top pair invariant mass distribution: A
  Window on new physics}},  {\em JHEP} {\bf 01} (2009) 047,
  [\href{http://arxiv.org/abs/0712.2355}{{\tt arXiv:0712.2355}}].

\bibitem{Djouadi:2019cbm}
A.~Djouadi, J.~Ellis, A.~Popov, and J.~Quevillon, {\it {Interference effects in
  $ t\overline{t} $ production at the LHC as a window on new physics}},  {\em
  JHEP} {\bf 03} (2019) 119, [\href{http://arxiv.org/abs/1901.03417}{{\tt
  arXiv:1901.03417}}].

\bibitem{Carena:2016npr}
M.~Carena and Z.~Liu, {\it {Challenges and opportunities for heavy scalar
  searches in the $ t\overline{t} $ channel at the LHC}},  {\em JHEP} {\bf 11}
  (2016) 159, [\href{http://arxiv.org/abs/1608.07282}{{\tt arXiv:1608.07282}}].

\bibitem{ATLAS:2024vxm}
{\bf ATLAS} Collaboration, G.~Aad et~al., {\it {Search for heavy neutral Higgs
  bosons decaying into a top quark pair in 140 fb$^{-1}$ of proton-proton
  collision data at $ \sqrt{s} $ = 13 TeV with the ATLAS detector}},  {\em
  JHEP} {\bf 08} (2024) 013, [\href{http://arxiv.org/abs/2404.18986}{{\tt
  arXiv:2404.18986}}].

\bibitem{CMS:2025dzq}
{\bf CMS} Collaboration, A.~Hayrapetyan et~al., {\it {Search for heavy
  pseudoscalar and scalar bosons decaying to a top quark pair in
  proton{\textendash}proton collisions at $\sqrt{s} = 13\,\textrm{TeV}$}},
  {\em Rept. Prog. Phys.} {\bf 88} (2025), no.~12 127801,
  [\href{http://arxiv.org/abs/2507.05119}{{\tt arXiv:2507.05119}}].

\bibitem{CMS:2020zti}
{\bf CMS} Collaboration, A.~Tumasyan et~al., {\it {Search for resonant and
  nonresonant production of pairs of dijet resonances in proton-proton
  collisions at $ \sqrt{s} $ = 13 TeV}},  {\em JHEP} {\bf 07} (2023) 161,
  [\href{http://arxiv.org/abs/2206.09997}{{\tt arXiv:2206.09997}}]. [Erratum:
  JHEP 25, 113 (2020)].

\bibitem{CMS:2019gwf}
{\bf CMS} Collaboration, A.~M. Sirunyan et~al., {\it {Search for high mass
  dijet resonances with a new background prediction method in proton-proton
  collisions at $\sqrt{s} =$ 13 TeV}},  {\em JHEP} {\bf 05} (2020) 033,
  [\href{http://arxiv.org/abs/1911.03947}{{\tt arXiv:1911.03947}}].

\bibitem{Drnevich:2024vfj}
M.~Drnevich, S.~Jiggins, J.~Katzy, and K.~Cranmer, {\it {Neural
  quasiprobabilistic likelihood ratio estimation with negatively weighted
  data}},  {\em Mach. Learn. Sci. Tech.} {\bf 6} (2025), no.~4 045023,
  [\href{http://arxiv.org/abs/2410.10216}{{\tt arXiv:2410.10216}}].

\bibitem{ATLAS:2025kmo}
{\bf ATLAS} Collaboration, G.~Aad et~al., {\it {Search for ttbar resonances in
  final states with exactly one or two leptons using 140 fb$^{-1}$ of pp
  collision data at $\sqrt{s}=13$ TeV with the ATLAS experiment}},
  \href{http://arxiv.org/abs/2512.17856}{{\tt arXiv:2512.17856}}.

\bibitem{BessidskaiaBylund:2016jvp}
O.~Bessidskaia~Bylund, F.~Maltoni, I.~Tsinikos, E.~Vryonidou, and C.~Zhang,
  {\it {Probing top quark neutral couplings in the Standard Model Effective
  Field Theory at NLO in QCD}},  {\em JHEP} {\bf 05} (2016) 052,
  [\href{http://arxiv.org/abs/1601.08193}{{\tt arXiv:1601.08193}}].

\bibitem{Englert:2019rga}
C.~Englert, P.~Galler, and C.~D. White, {\it {Effective field theory and scalar
  extensions of the top quark sector}},  {\em Phys. Rev. D} {\bf 101} (2020),
  no.~3 035035, [\href{http://arxiv.org/abs/1908.05588}{{\tt
  arXiv:1908.05588}}].

\bibitem{Georgi:1977gs}
H.~M. Georgi, S.~L. Glashow, M.~E. Machacek, and D.~V. Nanopoulos, {\it {Higgs
  Bosons from Two Gluon Annihilation in Proton Proton Collisions}},  {\em Phys.
  Rev. Lett.} {\bf 40} (1978) 692.

\bibitem{Djouadi:2005gi}
A.~Djouadi, {\it {The Anatomy of electro-weak symmetry breaking. I: The Higgs
  boson in the standard model}},  {\em Phys. Rept.} {\bf 457} (2008) 1--216,
  [\href{http://arxiv.org/abs/hep-ph/0503172}{{\tt hep-ph/0503172}}].

\bibitem{Plehn:2009nd}
T.~Plehn, {\it {Lectures on LHC Physics}},  {\em Lect. Notes Phys.} {\bf 844}
  (2012) 1--193, [\href{http://arxiv.org/abs/0910.4182}{{\tt
  arXiv:0910.4182}}].

\bibitem{Goria:2011wa}
S.~Goria, G.~Passarino, and D.~Rosco, {\it {The Higgs Boson Lineshape}},  {\em
  Nucl. Phys. B} {\bf 864} (2012) 530--579,
  [\href{http://arxiv.org/abs/1112.5517}{{\tt arXiv:1112.5517}}].

\bibitem{Seymour:1995np}
M.~H. Seymour, {\it {The Higgs boson line shape and perturbative unitarity}},
  {\em Phys. Lett. B} {\bf 354} (1995) 409--414,
  [\href{http://arxiv.org/abs/hep-ph/9505211}{{\tt hep-ph/9505211}}].

\bibitem{Papavassiliou:1996zn}
J.~Papavassiliou and A.~Pilaftsis, {\it {Gauge invariant resummation formalism
  for two point correlation functions}},  {\em Phys. Rev. D} {\bf 54} (1996)
  5315--5335, [\href{http://arxiv.org/abs/hep-ph/9605385}{{\tt
  hep-ph/9605385}}].

\bibitem{Papavassiliou:1997fn}
J.~Papavassiliou and A.~Pilaftsis, {\it {Effective charge of the Higgs boson}},
   {\em Phys. Rev. Lett.} {\bf 80} (1998) 2785--2788,
  [\href{http://arxiv.org/abs/hep-ph/9710380}{{\tt hep-ph/9710380}}].

\bibitem{Englert:2015zra}
C.~Englert, I.~Low, and M.~Spannowsky, {\it {On-shell interference effects in
  Higgs boson final states}},  {\em Phys. Rev. D} {\bf 91} (2015), no.~7
  074029, [\href{http://arxiv.org/abs/1502.04678}{{\tt arXiv:1502.04678}}].

\bibitem{Murayama:1992gi}
H.~Murayama, I.~Watanabe, and K.~Hagiwara, {\it {HELAS: HELicity amplitude
  subroutines for Feynman diagram evaluations}}, .

\bibitem{Degrande:2011ua}
C.~Degrande, C.~Duhr, B.~Fuks, D.~Grellscheid, O.~Mattelaer, and T.~Reiter,
  {\it {UFO - The Universal FeynRules Output}},  {\em Comput. Phys. Commun.}
  {\bf 183} (2012) 1201--1214, [\href{http://arxiv.org/abs/1108.2040}{{\tt
  arXiv:1108.2040}}].

\bibitem{Darme:2023jdn}
L.~Darm{\'e} et~al., {\it {UFO 2.0: the {\textquoteleft}Universal Feynman
  Output{\textquoteright} format}},  {\em Eur. Phys. J. C} {\bf 83} (2023),
  no.~7 631, [\href{http://arxiv.org/abs/2304.09883}{{\tt arXiv:2304.09883}}].

\bibitem{Alwall:2014hca}
J.~Alwall, R.~Frederix, S.~Frixione, V.~Hirschi, F.~Maltoni, O.~Mattelaer,
  H.~S. Shao, T.~Stelzer, P.~Torrielli, and M.~Zaro, {\it {The automated
  computation of tree-level and next-to-leading order differential cross
  sections, and their matching to parton shower simulations}},  {\em JHEP} {\bf
  07} (2014) 079, [\href{http://arxiv.org/abs/1405.0301}{{\tt
  arXiv:1405.0301}}].

\bibitem{Kniehl:1995tn}
B.~A. Kniehl and M.~Spira, {\it {Low-energy theorems in Higgs physics}},  {\em
  Z. Phys. C} {\bf 69} (1995) 77--88,
  [\href{http://arxiv.org/abs/hep-ph/9505225}{{\tt hep-ph/9505225}}].

\bibitem{ATLAS:2016bac}
{\bf ATLAS} Collaboration, M.~Aaboud et~al., {\it {Measurements of top quark
  spin observables in $ t\overline{t} $ events using dilepton final states in $
  \sqrt{s}=8 $ TeV pp collisions with the ATLAS detector}},  {\em JHEP} {\bf
  03} (2017) 113, [\href{http://arxiv.org/abs/1612.07004}{{\tt
  arXiv:1612.07004}}].

\bibitem{ATLAS:2022mlu}
{\bf ATLAS} Collaboration, G.~Aad et~al., {\it {Differential $ t\overline{t} $
  cross-section measurements using boosted top quarks in the all-hadronic final
  state with 139 fb$^{-1}$ of ATLAS data}},  {\em JHEP} {\bf 04} (2023) 080,
  [\href{http://arxiv.org/abs/2205.02817}{{\tt arXiv:2205.02817}}].

\bibitem{Buckley:2015lku}
A.~Buckley, C.~Englert, J.~Ferrando, D.~J. Miller, L.~Moore, M.~Russell, and
  C.~D. White, {\it {Constraining top quark effective theory in the LHC Run II
  era}},  {\em JHEP} {\bf 04} (2016) 015,
  [\href{http://arxiv.org/abs/1512.03360}{{\tt arXiv:1512.03360}}].

\bibitem{Brivio:2019ius}
I.~Brivio, S.~Bruggisser, F.~Maltoni, R.~Moutafis, T.~Plehn, E.~Vryonidou,
  S.~Westhoff, and C.~Zhang, {\it {O new physics, where art thou? A global
  search in the top sector}},  {\em JHEP} {\bf 02} (2020) 131,
  [\href{http://arxiv.org/abs/1910.03606}{{\tt arXiv:1910.03606}}].

\bibitem{forth}
D.~A.~B. Moreno, C.~Englert, and Y.~Peters, {\it {in preparation}}, .

\bibitem{Paszke:2019xhz}
A.~Paszke et~al., {\it {PyTorch: An Imperative Style, High-Performance Deep
  Learning Library}},  \href{http://arxiv.org/abs/1912.01703}{{\tt
  arXiv:1912.01703}}.

\bibitem{Pedregosa:2011ork}
F.~Pedregosa et~al., {\it {Scikit-learn: Machine Learning in Python}},  {\em J.
  Machine Learning Res.} {\bf 12} (2011) 2825--2830,
  [\href{http://arxiv.org/abs/1201.0490}{{\tt arXiv:1201.0490}}].

\bibitem{pytorchl}
W.~Falcon et~al., {\it Pytorchlightning/pytorch-lightning: 0.7.6 release},
  May, 2020.

\bibitem{Bahl:2025you}
H.~Bahl, R.~Kumar, and G.~Weiglein, {\it {Impact of interference effects on
  Higgs-boson searches in the di-top final state at the LHC}},  {\em JHEP} {\bf
  05} (2025) 098, [\href{http://arxiv.org/abs/2503.02705}{{\tt
  arXiv:2503.02705}}].

\bibitem{Robens:2026spl}
T.~Robens, {\it {Interference effects in new physics searches}},
  \href{http://arxiv.org/abs/2602.00256}{{\tt arXiv:2602.00256}}.

\bibitem{diego_baron_2026_19629894}
D.~Baron, {\it diegobaronm/ttbarresonancestudies: v1.0 - deep hunting paper
  version.},  Apr., 2026.

\bibitem{diego_baron_2026_19629945}
D.~Baron, {\it diegobaronm/peakdeepmaster: v1.0 - deep hunting paper version.},
   Apr., 2026.

\end{thebibliography}\endgroup
%%%%%%%%%%%%%%%%%%%
\end{document}